\documentclass[11pt,prc,aps,a4paper,groupedaddress,superscriptaddress,nofootinbib,showpacs
,preprintnumbers,onecolumn]{revtex4-1}
\usepackage{graphicx}
\usepackage{amsfonts}
\usepackage{amssymb,ulem,color}
\usepackage{natbib}
\usepackage{dcolumn}
\usepackage{bm}
\usepackage{subeqnarray}
\newcommand{\bwt}{\begin{widetext}}
\newcommand{\ewt}{\end{widetext}}
\newcommand{\beq}{\begin{equation}}
\newcommand{\eeq}{\end{equation}}
\newcommand{\bea}{\begin{eqnarray}}
\newcommand{\eea}{\end{eqnarray}}
\begin{document}
\title{Landau parameters for asymmetric nuclear matter with a strong magnetic field}
\author{M. \'Angeles P\'erez-Garc\'{\i}a}
\email{mperezga@usal.es}
\affiliation{Departamento de F\'{\i}sica Fundamental and IUFFyM, Universidad de Salamanca, E-37008 Salamanca}
\author{C.~Provid\^encia}
\email{cp@teor.fis.uc.pt}
\affiliation{Centro de F\' {\i}sica Computacional, Department of Physics, University of Coimbra, 3004-516 Coimbra, Portugal} 
\author{A. Rabhi}
\email{rabhi@teor.fis.uc.pt}
\affiliation{Centro de F\' {\i}sica Computacional, Department of Physics, University of Coimbra, 3004-516 Coimbra, Portugal} 
\affiliation{Laboratoire de Physique de la Mati\`ere Condens\'ee,
Facult\'e des Sciences de Tunis, Campus Universitaire, Le Belv\'ed\`ere-1060, Tunisia}

\date{\today}
\begin{abstract}
The Landau Fermi Liquid parameters are calculated for charge neutral asymmetric nuclear matter in beta equilibrium at zero temperature in the presence of a very strong magnetic field  with relativistic mean-field models. Due to the isospin structure of the system, with different populations of protons  and neutrons and spin alignment  to the field, we find non-vanishing  Landau mixing parameters. The existence of quantized Landau levels for the charged sector has some impact on the Landau parameters with the presence of discretized features in those involving the proton sector.  Using the Fermi liquid formalism singlet and triplet excited quasiparticle states are analyzed, and we find that in-medium effects and magnetic fields are competing, however, the former are more important in the interaction energy range considered. It is found that  for magnetic field strengths Log$_{10}$ B (G) $\le 17$ the relative low polarization of the system produces mild changes in the generalized Landau parameters with respect to the unmagnetized case, while for larger strengths there is a resolution of the degeneracy of the interaction energies of quasiparticles in the system.
\end{abstract}
\maketitle

\section{Introduction}
Asymmetric nuclear matter is presently an important topic proposed for study in  experiments at radioactive beam facilities such as FAIR \cite{fair} at GSI, SPIRAL2 \cite{ganil} at GANIL, ISAC-III at TRIUMF \cite{isac} or FRIB \cite{frib} at MSU, among others \cite{Blumenfeld}. These will allow the investigation of nuclei in regions of the nuclear chart far from the stability line. These nuclear regions, where the isospin asymmetry ratio, given by the ratio of the proton vector  number density, $n^v_p$, to baryonic vector number density, $n_B$, defined as $Y_P=n^v_p/n_B$,  largely departs from the $0.5$ value, are of interest in the study of stability of exotic nuclei as in the neutron rich nuclei. Recently studied examples of those are the isotopes of Ca \cite{z20} with $Z=20$ or the isotones $N=30$ of Ca and Sc with CLARA experiment \cite{clara} in Legnaro. The study of observables related to nuclei far from the isospin stability line allows for an improved description of neutron rich environments as those of interest in astrophysical scenarios of neutron star matter.  They are relevant to interiors of compact stellar objects like neutron stars (NS) arising in the aftermath of a supernova event. In this context, the rapid deleptonization of the NS following the gravitational collapse of the inner regions with the proton-electron capture process makes matter more neutron rich since (anti) neutrinos diffuse out of the star \cite{horowitz1, aziz10}. Also supernova matter can be considered to be constitued by a set of nuclei, in the nuclear statistical equilibrium (NSE) approximation, and presents a distribution of nuclei shifted to the $N>Z$ region \cite{nse, horowitz2}.

On Earth, the high temperature ($T$) and  small baryonic chemical potential ($\mu_B$)  region of matter phase space has been somewhat tested \cite{rhic} at  RHIC, and it will be possible to further test with other heavy-ion experiments like  ALICE \cite{alice} at
CERN. Although this improvement of our knowledge of the phases of  matter is certainly
valuable, the phase diagram of nuclear matter relevant for the equation of state (EOS) of NS is that of {\it cold} asymmetric nuclear matter in the low temperature range.

Historically, most of the existing literature on the nuclear matter EOS characterization \cite{revieweos} has neglected the input from external fields, in particular, in the case of magnetic fields, this is claimed due to the tiny value of the nuclear magnetic moment \cite{pdb}. On terrestrial experiments recent estimations of the magnetic field strength that could be produced dynamically at CERN  or BNL energies \cite{skokov} are of the order $B \approx 10^{17}-10^{19}$ G. In nature, we have another indication of  sources of intense magnetic fields of astrophysical origin such as magnetars \cite{glen00}.  Magnetars are neutron stars which may have {\it surface} magnetic fields $B \approx 10^{15}$ G \cite{duncan,usov,pacz} discovered in the  X-ray and $\gamma$-ray electromagnetic spectrum (for a review see \cite{harding06}). They are identified with the anomalous X-ray pulsars (AXP) and soft $\gamma$-ray repeaters \cite{sgr}. 

Taking as reference the critical field, $B^e_c$, at which the electron cyclotron energy is equal
to the electron mass, $B^e_c=4.414 \times 10^{13}$ G, we define $B^*=B/B^e_c$.  It has been shown
by several authors \cite{chakrabarty96,broderick, aziz08, aziz09} that  magnetic fields larger than $B^*=10^5$  will affect the EOS of
compact stars. In particular, field-theoretical descriptions based on several parametrizations
of the non-linear Walecka model (NLWM) \cite{bb}  show an overall similar behavior. According to the scalar
virial theorem \cite{virial} the {\it interior} magnetic field strength could be as large as
$B=10^{18}$ G so, in principle, this is the maximum field strength that is meaningful to consider. 

In 1959 the formal theory for treatment of low temperature (non-superfluid)  fermion systems, known as normal {\it  Fermi Liquids},  was developed by Landau \cite{landau1,landau2} to describe the behavior of $^3$He below $100$ mK. With this Fermi Liquid Theory (FLT) (see a recent reference \cite{FLT}) the  excited states in the system could be described as  quasiparticles (qp) as long as these states have sufficiently long lifetimes. At low temperature, the small excitation energy (compared to the chemical potential) will assure this fact. In the context of the FLT, the so-called {\it  Landau parameters}, can parametrize the interaction energy between a pair of qp in the medium. Previous works have attempted to partially study the behavior of Landau parameters for non-magnetized symmetric nuclear matter or neutron matter  \cite{caillon1,caillon2,matsui} or in magnetized matter under the presence of a magnetic field either in a non-relativistic formalism \cite{ang1,ang2, ang3} or in  magnetized matter without considering B field including exchange terms in a relativistic way \cite{nino}.

In this work we will be interested in calculating the Landau Fermi Liquid parameters for an isospin asymmetric nuclear system in beta equilibrium and in charge neutrality under the effect of an intense magnetic field. The FLT used in this case must describe relativistically the more  general condition of a magnetized non-pure isospin system to account for the fact that the intense magnetic field can modify isospin populations and partially align nucleon magnetic moments with respect to the case of vanishing magnetic field. In addition, the different dynamics of proton and neutron sectors under the presence of a magnetic field (including the existence of anomalous nucleon magnetic moments) will have effects in the  Landau parameter computation showing discretized or continue features for protons and neutrons, respectively. 
In section \ref{formalism} we introduce the relativistic lagrangian model used in this work and
the  generalized formalism of the FLT for charge neutral isospin asymmetric hadronic systems under the presence of a
magnetic field. In section \ref{landaucoef} we discuss the explicit form of the matricial
structure of the coefficients describing the interaction of qp in the magnetized system through
the Landau parameters. In section \ref{results} we analyze the obtained coefficients  for either individual spin quantum numbers or total spin (singlet or triplet) for  the qp excitations for the electrically neutral system
configurations calculated under beta equilibrium and the Landau
parameter behavior in presence of a strong magnetic field and, finally, in section \ref{summary}, we summarize and draw some conclusions.

\newpage
\section{The formalism}
\label{formalism}
For the description of the EOS of neutron star matter, we employ a relativistic field-theoretical approach in which the baryons, neutrons (n) and protons (p), interact via the exchange of $\sigma-\omega-\rho$ mesons in the presence of a uniform magnetic field $B$ 
along the $z$-axis. The Lagrangian density for the  TM1 parametrization \cite{tm1} of the non-linear Walecka model (NLWM) reads \cite{bb}
\beq
{\cal L}= \sum_{b=n, p}{\cal L}_{b} + \sum_{m=\sigma,\omega,\rho}{\cal L}_{m}+ \sum_{l=e}{\cal L}_{l}.
\label{lan}
\eeq
The baryon ($b$=$n$, $p$), meson ($m=\sigma$, $\omega$ and  $\rho$) and lepton ($l=e$) lagrangians are given by ($c=\hbar=$1),
\beq
{\cal L}_{b}=\bar{\Psi}_{b}\left(i\gamma_{\mu}\partial^{\mu}-q_{b}\gamma_{\mu}A^{\mu}- 
m_{b}+g_{\sigma}\sigma
-g_{\omega}\gamma_{\mu}\omega^{\mu}-\frac{1}{2}g_{\rho}\tau_{3 b}\gamma_{\mu}\rho^{\mu}
-\frac{1}{2}\mu_{N}\kappa_{b}\sigma_{\mu \nu} F^{\mu \nu}\right )\Psi_{b},
\eeq
\bea
{\cal L}_{m}=\frac{1}{2}\partial_{\mu}\sigma \partial^{\mu}\sigma
-\frac{1}{2}m^{2}_{\sigma}\sigma^{2}-\frac{1}{3!}\kappa \sigma^{3} -\frac{1}{4!}\lambda \sigma^{4}
+\frac{1}{2}m^{2}_{\omega}\omega_{\mu}\omega^{\mu}+\frac{1}{4!}\xi g^4_{\omega}(\omega_{\mu}\omega^{\mu} )^2-\frac{1}{4}\Omega^{\mu \nu} \Omega_{\mu \nu}  \nonumber \\
-\frac{1}{4} F^{\mu \nu}F_{\mu \nu}+\frac{1}{2}m^{2}_{\rho}\rho_{\mu}\rho^{\mu}-\frac{1}{4}  P^{\mu \nu}P_{\mu \nu},
\eea
\beq
{\cal L}_{l}=\bar{\Psi}_{l}\left(i\gamma_{\mu}\partial^{\mu}-q_{l}\gamma_{\mu}A^{\mu}- m_{l}\right )\Psi_{l},
\label{lagran_l}
\eeq
where $\Psi_{b}$ , $\Psi_{l}$ are the baryon  and lepton Dirac fields respectively. 

The nucleon isospin $z$-projection for the proton (neutron) is denoted by $\tau_{3 p}=1$ ( $\tau_{3n}=-1$). The nucleon mass is $m_{b}$ ($m_b=m_n=m_p=938$ MeV), its charge is $q_b$ and the baryonic  anomalous magnetic moments (AMM) are introduced via the coupling to the electromagnetic field tensor with
$\sigma_{\mu \nu}=\frac{i}{2}\left[\gamma_{\mu},  \gamma_{\nu}\right] $ and strength
$\kappa_{b}$. In particular, $\kappa_{n}=-1.91315$ for the neutron and $\kappa_{p}=1.79285$ for the
proton.  $m_l$ and $q_l$ are the mass and charge of the lepton. We will consider
the simplest model where the leptonic sector is formed just by electrons ($l=e$), with no anomalous magnetic
moment, providing charge neutrality in this astrophysical
scenario. Despite there is a non-zero electron AMM its value  \cite{pdb} is tiny when compared to that in the hadronic sector and it was shown that this contribution is negligible for the magnetic fields of interest in astrophysics if properly introduced \cite{duncan00}. 
The mesonic and electromagnetic field strength tensors are given by their usual expressions: $\Omega_{\mu \nu}=\partial_{\mu}\omega_{\nu}-\partial_{\nu}\omega_{\mu}$, $P_{\mu 
\nu}=\partial_{\mu}\rho_{\nu}-\partial_{\nu}\rho_{\mu}$, and  $F_{\mu
\nu}=\partial_{\mu}A_{\nu}-\partial_{\nu}A_{\mu}$.

The electromagnetic field is assumed to be externally generated (and thus
has no associated field equation), and only frozen-field configurations will be considered in this work. To calculate the thermodynamic conditions in this charge neutral asymmetric nuclear system in beta equilibrium with an intense magnetic field,  isoscalar and isovector current conservation must be imposed. Explicitly, the following conditions are fulfilled: i) electrical charge neutrality ii) conservation of baryonic charge iii) mesonic field equations selfconsistency. In addition, we will assume thorought this work that neutrinos scape freely and therefore there is no neutrino trapping.

The field equations of motion are determined from the Euler-Lagrange equations arising from the lagrangian density in Eq.(\ref{lan}). Under the conditions of the present calculation, a relativistic mean field (RMF) approximation will be used so that the space-time varying fields are replaced by a homogeneous value, $\phi(x_{\mu}) \rightarrow \phi$. In this way, the  Dirac equation for a nucleon is given by,
\bea
(i\gamma_{\mu}\partial^{\mu}-q_{b}\gamma_{\mu}A^{\mu}-(m_{b}
-g_{\sigma}\sigma)-g_{\omega}\gamma_{\mu}\omega^{\mu} \cr
-\frac{1}{2}g_{\rho}\tau_{3 b}\gamma_{\mu}\rho^{\mu} 
-\frac{1}{2}\mu_{N}\kappa_{b}\sigma_{\mu \nu} F^{\mu \nu}) \Psi_{b}&=&0 ,
\label{MFbary}
\eea
where the effective baryon mass is $m^{*}=m_{b}-g_{\sigma}\sigma$. For leptons,
\beq
\left(i\gamma_{\mu}\partial^{\mu}-q_{l}\gamma_{\mu}A^{\mu}-m_{l}\right) \Psi_l=0.
\label{MFlep}
\eeq
For meson fields we obtain,
\bea
m^{2}_{\sigma} \sigma  + \frac{1}{2}\kappa\sigma^2+ \frac{1}{3!}\lambda\sigma^3&=&g_{\sigma}\left(n^{s}_{p}+n^{s}_{n}\right) 
=g_{\sigma}n^{s}\label{mes1}, \\
m^{2}_{\omega} \omega^{0} + \frac{1}{3!}\xi g_{\omega}^4 \left({\omega^0}^2-\boldsymbol{\omega}^2\right){\omega^0} &=& g_{\omega}\left(n^v_{p}+n^v_{n}\right)= 
g_{\omega}n_{B}\label{mes2}, \\
m^{2}_{\omega}\boldsymbol{\omega} + \frac{1}{3!}\xi
g_\omega^4\left({\omega^0}^2-\boldsymbol{\omega}^2\right)\boldsymbol{\omega} &=& g_{\omega}\boldsymbol{j}_{B}\label{mes2v}, \\
m^{2}_{\rho} \rho^{0}  &=& \frac{1}{2}g_{\rho}\left(n^v_{p}-n^v_{n}\right) =\frac{1}
{2}g_{\rho}n_{3}\label{mes3},\\
m^{2}_{\rho} \boldsymbol{\rho}  &=& 
\frac{1}{2}g_{\rho}\boldsymbol{j}_{3}
\label{mes3v},
\eea
where we use the notation as in the work of Matsui \cite{matsui} and $n_B$ is the baryonic
(vector) particle number density constructed as the sum of (vector) particle number density of protons ($n^v_p$) and
neutrons ($n^v_n$), $n_B =n^v_p +n^v_n$ . 

The baryon current $\boldsymbol{j_B}=\boldsymbol{j_p}+\boldsymbol{j_n}$  is also the sum of the
proton (${\mathbf j_p}$) and neutron (${\mathbf j_n}$) currents. $n^s=n^s_p+n^s_n$ is the scalar density constructed from that of protons ($n^s_p$) and neutrons ($n^s_n$). $n_3=n^v_p-n^v_n$ and
$\boldsymbol{j_3}=\boldsymbol{j_p}-\boldsymbol{j_n}$ are the isoscalar particle number density and isovector baryon current, respectively. When solving for equilibrium conditions in the nuclear system governed by Eqs.(\ref{MFbary})-(\ref{mes3v}) we impose ${\bf j_B=0}$ and ${\bf j_3=0}$. Then, we have for  the $\sigma$ field, 
\beq
g_{\sigma} \sigma=\frac{g^2_{\sigma}}{ {m'}^2_{\sigma}}\, 
n^s,
\eeq
with ${m'}^2_{\sigma}=m_{\sigma}^2 + \frac{1}{2}\kappa\sigma+ \frac{1}{3!}\lambda\sigma^2$. For the $\omega^0$ field we get
\beq
g_{\omega}  {\omega^0}=\frac{g_{\omega}^2}{{m'}^2_\omega}n_{B},
\eeq
with ${m'}_\omega^2=m_{\omega}^2 + \frac{1}{3!}\xi g_{\omega}^4{\omega^0}^2$.

When the Dirac equation for nucleons Eq.(\ref{MFbary}) is solved, a magnetic field B in the $z$-direction given by $\boldsymbol{B}=B\boldsymbol{ \hat{k}}$ is used. The energies for the quasi-protons and quasi-neutrons in the medium with spin $z$-projection, $s$,  are given by the following expressions~\cite{broderick},
\bea
\epsilon^{p}_{\nu, s}&=& \sqrt{k^{2}_{z}+\left(\sqrt{m^{* 2}_{p}+2\nu q_{p}B}-s\mu_{N}\kappa_{p}B \right) 
^{2}}+g_{\omega} \omega^{0}+\frac{1}{2}g_{\rho}\rho^{0}, \label{enspc1}\\
\epsilon^{n}_{s}&=& \sqrt{k^{2}_{z}+\left(\sqrt{m^{* 2}_{n}+k^{2}_{\perp}}-s\mu_{N}\kappa_{n}B 
\right)^{2}}+g_{\omega} \omega^{0}-\frac{1}{2}g_{\rho}\rho^{0}\label{enspc2},
\eea
where $\nu=n+\frac{1}{2}-sgn(q_b)\frac{s}{2}=0, 1, 2, \ldots$ enumerates the quantized  Landau levels for  protons with electric 
charge $q_p$. The quantum number $s$ is $+1$ for spin up and $-1$ for spin down quasiparticles. Due to
the fact that the magnetic field is taken in the $z$-direction, it is useful to define three-momentum (${\mathbf k}$) components along parallel ($k_z$) and perpendicular (${\bf k_{\perp}}$) directions. Then, for neutrons, the surface of constant energy is an ellipsoid while, for
protons,  constant energy surfaces are formed by circumferences on nested cylinders with radius labeled by the Landau level, see next section. In this work we are mainly interested in hadronic properties, and electron dynamics will be such that at high B field strengths, they will mostly be in the low Landau levels. For completeness we write the expressions of the scalar and vector densities for protons and neutrons for both spin polarizations as follows~\cite{broderick}
\bea
n^{s}_{p}&=&\frac{q_{p}Bm^{*}_{p}}{2\pi^{2}}\sum_{\nu=0}^{\nu_{\mbox{\small max}}}\sum_{s}\frac{\sqrt{m^{* 2}_{p}+2\nu 
q_{p}B}-s\mu_{N}\kappa_{p}B}{\sqrt{m^{* 2}_{p}+2\nu q_{p}B}}\ln\left|\frac{k^{p}_{F,\nu,s}+E^{p}_{F}}
{\sqrt{m^{* 2}_{p}+2\nu q_{p}B}-s\mu_{N}\kappa_{p}B} \right|, \cr
n^{s}_{n}&=&\frac{m^{*}_{n}}{4\pi^{2}}\sum_{s} \left[E^ {n}_{F}k^{n}_{F, s}-\bar{m}^{2}_{n}\ln\left|
\frac{k^{n}_{F,s}+E^{n}_{F}}{\bar{m}_{n}} \right|\right],  \cr
n^{v}_{p}&=&\frac{q_{p}B}{2\pi^{2}}\sum_{\nu=0}^{\nu_{\mbox{\small max}}}\sum_{s}k^{p}_{F,\nu,s},  \cr
n^{v}_{n}&=&\frac{1}{2\pi^{2}}\sum_{s}\left[ \frac{1}{3}\left(k^{n}_{F, s}\right) ^{3}-\frac{1}
{2}s\mu_{N}\kappa_{n}B\left(\bar{m}_{n}k^{n}_{F,s}+E^{n 2}_{F}\left(\arcsin\left( \frac{\bar{m}_{n}}
{E^{n}_{F}}\right) -\frac{\pi}{2} \right)  \right) \right] ,
\eea
where $k^{p}_{F, \nu, s}$, $ k^{n}_{F, s}$ are the Fermi momenta of protons and neutrons related to the proton and neutron Fermi energies, $E^{p}_{F}$ and $E^{n}_{F}$, respectively, by
\bea
k^{p 2}_{F,\nu,s}&=&E^{p 2}_{F}-\left[\sqrt{m^{* 2}_{p}+2\nu q_{p}B}-s\mu_{N}\kappa_{p}B\right] ^{2}, \cr
k^{n 2}_{F,s}&=&E^{n 2}_{F}-\bar{m}^{2}_{n}, 
\eea
and $\bar{m}_{n}=m^{*}_{n}-s\mu_{N}\kappa_{n}B\equiv \bar m_n^s$. The summation in $\nu$ in the above expressions terminates at $\nu_{max}$,
 the largest value of $\nu$ for which the square of Fermi momentum of  the charged particle is still positive and corresponds to the closest integer from below defined by the ratio
\beq
\nu_{max}=\left[\frac{(E^p_F+s\,\mu_N\,\kappa_p\,B)^2-{m_p^*}^2}{2 |q_p|\,  B}\right].
\eeq
 For electrons $ k^{e }_{F,\nu,s}=\sqrt{{E^{e}_{F}}^2-({m^{ 2}_{e}+2\nu q_{e}B})}$ and $\nu_{max}$ does not have the same value as that for protons.

\section{Landau Fermi Liquid parameters}
\label{landaucoef}
We now calculate the Landau parameters using a generalized formulation of Landau Theory of Fermi Liquids  \cite{FLT}, from the variations of the energy density of the system, $\cal {E}$,  given by \cite{caillon1}  
\bea
{\cal E}=\frac{1}{2} \frac{g^2_{\omega} m^2_{\omega}}{m^{' 4}_{\omega} } n^2_B-\frac{1}{2} \frac{g^2_{\omega} m^2_{\omega} }{m^{' 4}_{\omega} } \boldsymbol{j}^2_B+
\frac{1}{4!} \frac{  \xi g_{\omega}^8 }{m^{' 8}_{\omega} } n^4_B
+\frac{1}{4!} \frac{ \xi g_{\omega}^8 }{m^{' 8}_{\omega} } \boldsymbol{j}^4_B 
-\frac{2}{4!} \frac{\xi g_{\omega}^8 }{m^{' 8}_{\omega} } \boldsymbol{j}^2_B n^2_B \nonumber\\
+\frac{1}{8} \frac{g^2_{\rho}}{m^2_{\rho}} n^2_3-\frac{1}{8} \frac{g^2_{\rho}}{m^2_{\rho}} \boldsymbol{j}^2_3+\sum_{i,\nu,s} n^p_{i} E^{p}_{i \nu, s} +\sum_{i,s} n^n_{i} E^{n}_{i s}+\sum_{i,\nu, s} n^l_{i} E^{l}_{i \nu, s} \nonumber \\
+ \frac{1}{2} \frac{m^2_{\sigma}}{g^2_{\sigma}} (m_{b}-m^{*})^2+\frac{1}{3!}\frac{\kappa}{g^2_{\sigma}} (m_{b}-m^{*})^{3} +\frac{1}{4!} \frac{\lambda}{g^2_{\sigma}} {(m_{b}- m^{*})}^{4}+\frac{B^2}{2},
\eea
where we have defined $n^p_{i}=n^p(k_i,\nu, s_i)$ as the occupation number of the quasiprotons and $n^n_{i}=n^n(k_i,s_i)$ for the quasineutrons. For leptons the occupation number is $n^l_{i}=n^l(k_i,\nu, s_i)$. Also we have defined the energies,
\bea
E^{p}_{i\nu, s}&=& \sqrt{{\cal{K}}^{p 2}_{zi}+\left(\sqrt{m^{* 2}_{p}+2\nu q_{p}B}-s\mu_{N}\kappa_{p}B \right) 
^{2}}, \label{enspc1a}\\
E^{n}_{i s}&=& \sqrt{{\cal{K}}^{n 2}_{zi}+\left(\sqrt{m^{* 2}_{n}+{\cal K}^{2}_{\perp i}}-s\mu_{N}\kappa_{n}B 
\right)^{2}}, \label{enspc2a} 
\eea
and analogous for electrons, $E^{l}_{i \nu, s}$. We use generalized three-momenta depending on isospin
\beq
\boldsymbol{{\cal K}}^j_{i}= \boldsymbol{k}_{i} - \boldsymbol{{\cal V}}^j_{i}, \quad j=p,\, n
\eeq
with  $\boldsymbol{{\cal V}}^p=g_\omega \boldsymbol{\omega}+\frac{1}{2}g_\rho \boldsymbol{\rho}$ and  $ \boldsymbol{{\cal V}}^n=g_\omega \boldsymbol{\omega}-\frac{1}{2}g_\rho \boldsymbol{\rho}$. The equation for the nucleon effective mass can be written as
\beq
m^*=m_b-g_{\sigma} \sigma=m_b-\frac{g^2_{\sigma}}{ {m'}_{\sigma}^2}\, 
(n^s_p+n^s_n),
\eeq
where 
\beq
n^s_n= \sum_{i,s} \frac {n^n_{is} m^*}{E^n_{is}}
\left(1- \frac { s\mu_{N}\kappa_{n}B } {  \sqrt{m^{* 2}_{n}+{\cal K}^{2}_{\perp i}} }  \right),
\eeq
\beq
n^s_p=\sum_{i,\nu, s} \frac {n^p_{i,\nu,s} m^*}{E^p_{i \nu s}} \frac {\bar m^p_{\nu s}} {\bar m^p_{\nu s}+s\mu_{N}\kappa_{p}B },
\eeq
where the following definition  has been used
\beq
\bar m^p_{\nu s}=\sqrt{{m^*}^2 + 2 q_p \nu B}- s\kappa_p \mu_N B=\tilde \epsilon^p_{\nu s}- s\kappa_p \mu_N B.
\eeq
The vector current for quasiprotons is written as 
\beq
{\mathbf j}_p=\sum_{i,\nu, s} \frac{ \boldsymbol{{\cal K}}^p_{i} n^p_{i} }
{[ \boldsymbol{{\cal K}}^{2 p}_{i} +(\bar m^p_{\nu s})^{2} ]^{1/2}},
\label{jp}
\eeq
and for quasineutrons, 
\beq
{\mathbf j}_n=\sum_{i,s} n^n_{is} \left[ \frac {\boldsymbol{ {\cal K}}^n_{\perp i} } {E^n_{is}}
\left( 1- \frac { s\mu_{N}\kappa_{n}B } {  \sqrt{m^{* 2}_{n}+\boldsymbol{{\cal K}}^{n 2}_{\perp i}}}   \right)
+ 
\frac{\boldsymbol{ {\cal K}}^n_{z i}}{E^n_{is}} \right],
\label{jn}
\eeq
so that ${\mathbf j_B}$ and ${\mathbf j_3}$ can be constructed.

According to the FLT \cite{FLT} the first variation of the energy density of the system, $\cal{E}$, with respect to the ocupation number for qp with isospin of jth-type,  $n^j_i$, defines the qp energy, $\epsilon^j_i$. Let us notice that in reduced notation the index $i$ means $(i,\nu, s)$ for quasi-protons and $(i,s)$ for quasi-neutrons:
\beq
\delta {\cal E}=\sum_{i,j} \epsilon^j_i \delta n^j_i.
\label{deltaE}
\eeq
and
\beq
\epsilon_i^p=  E^{p}_{i, \nu, s} + \frac{g^2_\omega }{m^{'2}_\omega} n_B + \frac{g^2_{\rho}}{4 m^2_{\rho}} n_3,
\eeq
\beq
\epsilon_i^n= E^{n}_{i s} + \frac{g^2_\omega}{m^{'2}_\omega} n_B - \frac{g^2_{\rho}}{4 m^2_{\rho}} n_3.
\eeq
The single quasiparticle energy has, each, two explicit contributions, one  due to the motion under the influence of a strong quantizing magnetic fiel and another due to the motion in a medium with mesonic self-interacting fields. To calculate the Landau parameters we use the standard approach \cite{FLT} but generalizing to the case when there are external fields. These will allow to extract information on the interaction energies of quasiparticles of spin $z$-projection $(s, s')$ and isospin $(i,j)$  and, if they are protons, different Landau levels $(\nu,\nu')$ in the system. In a generalized system with a qp state the Landau parameters are calculated as the second derivative of the energy density of the system, ${\cal E}$, with respect to the qp state with occupation number $n^j_{l,s'}$, that is $j$-isospin, spin $s'$ and momentum ${\mathbf {\cal K}}^j_l$ (if they are quasiprotons they include the additional quantum number $\nu$, in the way $n^j_{l \nu s'}$). Using Eq. (\ref{deltaE}) it is defined, 
\beq
f^{ij}_{rlss'}=\frac{\partial \epsilon^{i}_{r, s}}{\partial n^j_{l,s'}}.
\eeq
From the original formulation by Landau of the pure neutron system without considering spin degrees calculated in \cite{landau1,landau2}
the above expression generalizes to a $(2\nu_{max}+2) \times (2\nu_{max}+2)$  matrix, $f_L$, in isospin and spin space. In this way we have a characterization of the nuclear system when an external quantizing magnetic field is considered,
\beq
f_L =
\left( {\begin{array}{cc}
 f^{pp}_{ilss'\nu\nu'} & f^{pn}_{ilss'\nu} \\
f^{np}_{ilss'\nu'} & f^{nn}_{ilss'}  \\
 \end{array} } \right).
\label{matrixf}
\eeq
The  detailed calculation of the matrix elements is given in the appendix in section \ref{appendix}. 

In order to obtain the relativistic Landau parameters  we must consider in the context of the FLT that the interactions of the effective quasiparticles in the system will take place close to the Fermi surfaces, since the lifetime of these excitations varies inversely with the departure of its energy, $E$,  from the Fermi energy $\tau \approx \frac{1}{(E-E_F)^2}$.

Since the matrix elements in Eq.(\ref{matrixf}) have dimensions of energy divided by number density it seems convenient to define new dimensionless coefficients by multiplying them by the density of states at each Fermi level for quasiprotons, $N_p^{s, \nu}$ and quasineutrons, $N_n^s$, at the Fermi surface with a given spin projection. In the case of protons the quantized level filling must be carefully considered and the definition of density of states at the Fermi level and in a given Landau level with polarization $s$ is
\begin{eqnarray}
N_p^{s,\nu}&=&\frac{1}{V}\sum_{\mathbf{k}} \delta(\epsilon_{k^p_z,\nu,s}-\mu)=\frac{|q_p|B}{2\pi \, L}\sum_{{k^p_z}}
\delta(\epsilon_{k^p_z,\nu,s}-\mu) \nonumber \\
&\to&\frac{|q_p|B}{2\pi^2}\int_0^{\infty} dk^p_z
\delta(\epsilon_{k^p_z,\nu,s}-\mu)=\frac{|q_p|B}{2\pi^2}\int_0^{\infty} d\epsilon_{k^p_z,\nu,s}
\frac{\epsilon_{k^p_z,\nu,s}}{k^p_z}  \delta(\epsilon_{k^p_z,\nu,s}-\mu)\nonumber \\
&=&\frac{|q_p|B}{2\pi^2}\frac{\mu}{k^p_{F,\nu,s}},
\end{eqnarray}
where $\mu=E^p_F$ is the proton Fermi energy.
Summing over all possible levels we have,
\beq
N_p^s=\left( \frac{\partial n^p_s}{\partial \epsilon^p_{s}}\right)_{E^p_F}= \sum_\nu N_p^{s, \nu}=\frac{q_p \, B}{2\pi^2} \sum_\nu \frac{_{E^p_{F}}}{k_{F,\nu,s}^{p}},
\label{den_np}
\eeq
where  $ k^p_{F, \nu, s}=\sqrt{{E^p_F}^2-\left({\bar{m}}^p_{\nu s}\right)^2}$. For neutrons,
\beq
N_n^s=\left( \frac{\partial n^n_s}{\partial \epsilon^n_{s}} \right)_{E^n_F}=
 \frac{E^n_F}{2\pi^2}
\left\{ k_{F,s}^{n}-s \mu_{N} \kappa_n B \left[\mbox{arcsin} \left(\frac{\bar m^s_n}{E_F^n}\right)-\frac{\pi}{2}\right] \right\}.
\label{den_nn}
\eeq

For the mixed states  and due to the fact that the $f^{ij}_{ss'}$ arise from two-body operators we can define the density,
\beq
N_{ij}^{ss'}=\sqrt{N_i^s N_j^{s'}}.
\label{den_nij}
\eeq
Notice that this is not the only possible definition for the mixed density of states, however,  the correct limit  for isospin pure systems  and $B=0$ limit can be recovered when used. Using this prescription, dimensionless coefficients $F^{ij}_{ss'}$ can be defined \cite{landau1,landau2,FLT}  so that:
\beq
F^{ss'}_{ij}=N_{ij}^{ss'} f^{ij}_{ss'}.
\label{Fij}
\eeq

\begin{figure}[hbtp]
\begin{center}
\includegraphics [angle=0,width=0.3\linewidth]{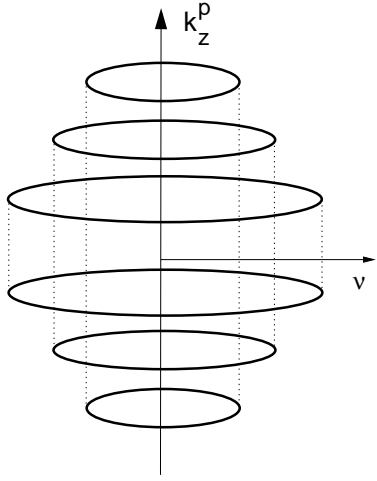}
\caption{Cilindrical  shape of constant energy for quasiprotons. For spin projection $s$ the Fermi surface are concentric circles with values $k^p_z= k^p_{F,\nu,s}$  and radius ${\bar{m}}^p_{\nu s}$. Notice that as $\nu$ grows $k^p_z$ is smaller.}
\label{Fig1}
\end{center}
\end{figure}

\begin{figure}[hbtp]
\begin{center}
\includegraphics [angle=0,width=0.3\linewidth]{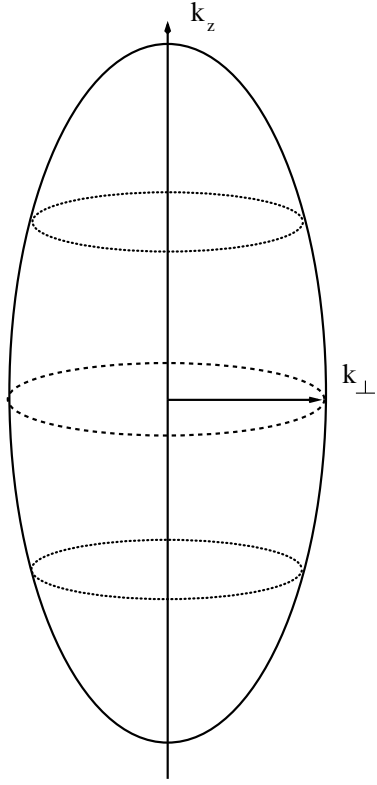} 
\caption{Ellipsoidal shape of constant energy for quasineutrons. For spin projection $s$ the Fermi surface has maximum values $|k|^{n, max}_{z, s}= k^n_{F,s}$  and  $|\boldsymbol{k}|^{n, max}_{\perp, s}=k^n_{F\perp, s}$}.
\label{Fig2}
\end{center}
\end{figure}

The Fermi surfaces are selected between the energy surfaces given by Eq.(\ref{enspc1a}) and Eq.(\ref{enspc2a}) in the system at equilibrium. For protons these are cilindrical holes with $k^p_{z}$ taking values $\pm k^p_{F\nu s}$ at the Landau level defined by $\nu$ for a given spin projection $s$ as shown in Fig.\ref{Fig1}.
 For each spin projection, $s=\pm1$ and a given $\nu$, the proton Fermi surface are two circles with $k^p_z=\pm
k^p_{F\nu s}$. For neutrons, instead, the Fermi surface for given $s$ is shown in Fig.\ref{Fig2} as an ellipsoid of maximum perpendicular momentum $|\boldsymbol{k}|^{n, max}_{\perp, s}=k^n_{F\perp, s}$. This maximum Fermi momentum in the transverse direction is
\beq
|\boldsymbol{k}|^{n, max}_{\perp, s}=k^n_{F\perp, s}=\sqrt{\left({E^n_F}+s\, \kappa_n \mu_N
B\right)^2-{m^*}^2},
\label{maxkp}
\eeq 
while in the $z$-direction the  $k^n_z$ value attains a maximum value,
\beq
 |k|^{n, max}_{z, s}=k^n_{Fs}=\sqrt{{E^n_F}^2-\left({\bar{m}}_n^{s}\right)^2}.
\eeq 

In regular FLT the relevant interaction takes place on the Fermi surfaces and then the Landau parameters depend on  the density, $n_B$, and the angle, $\theta$, between qp three momenta being possible to perform their expansion in Legendre polynomials, $P_l(cos(\theta))$. Taking averages over angular dependence of the Landau parameters shows that the only terms remaining are the $l=0$ terms. Then, for a pure isospin system the relations $f_l=\frac{f_{s, s}+f_{s.-s}}{2}$, and $g_l=\frac{f_{s, s}-f_{s.-s}}{2}$ hold and in this way values for $F_0=N_0 f_0$ and $G_0=N_0 g_0$ were the first to be historically obtained by Landau \cite{landau1,landau2} and can be obtained from indirect experimentally measured observables \cite{matsui}. 
On general grounds we can define auxiliar combinations of Landau parameters, 
\beq
F^{+}_{ij} =\frac{F^{++}_{ij}+F^{+-}_{ij}}{2}, \quad F^{-}_{ij} =\frac{F^{--}_{ij}+F^{-+}_{ij}}{2},
\eeq
\beq
G^{+}_{ij} =\frac{F^{++}_{ij}-F^{+-}_{ij}}{2} ,\quad  G^{-}_{ij} =\frac{F^{--}_{ij}-F^{-+}_{ij}}{2},
\eeq
and, from them, we define combinations with singlet $(S=0)$ or triplet $(S=1)$ qp states as,
\beq
F^{(S=0)}_{ij} =F^+_{ij}+ F^-_{ij},\quad  F^{(S=1)}_{ij} =G^+_{ij}+ G^-_{ij}.
\label{fs0s1}
\eeq
If the $B \to 0$ limit is taken, we recover from the previous expression the usual $F_0$ and $G_0$ in the normal FLT.

In our work, in a similar way, and in order to evaluate the angular averages of coefficients over qp Fermi surfaces we must take into account  the different dynamical behavior due to the qp electrical charge. For protons there are several Landau levels  that can be populated over cilindrical holes with $k^p_{z}=\pm k^p_{F\nu s}$. The average of a function ${\tilde f(k^p_{z},s,s' , \nu,\nu')}$ is performed as
\begin{equation}
<{\tilde f}>_{ss'}=\frac{\sum_{\nu,\nu'} \tilde f(k^p_{z},s,s' , \nu,\nu') N^{\nu s}_p N^{\nu' s'}_p}  {\sqrt{N^{s}_p N^{s'}_p}} .\label{def_f}
\end{equation}

This definition gives the correct $B\to 0$ limit as the one obtained with a regular FLT $B=0$ calculation. For neutrons the Fermi surface is an ellipsoid  defined by Eq.(\ref{enspc2a}) and the average
of a given function ${\tilde g(k^n_z,s)}$ should be performed using the fact it presents axial
symmetry. We will perform an integration over $k^n_\perp$ resulting from the projection of the Fermi volume over the
plane, that is a disk $\cal{ S_{\perp}}$  of radius $|\boldsymbol{k}|^{n, max}_{\perp, s}$ as given by Eq.(\ref{maxkp}). Then, in order to average we must   replace  the $k^n_z$ by its value on the Fermi surface 
\beq
k^n_z=\pm \sqrt{{E^n_F}^2-\left({\bar{m}}^n_{s}(k_\perp)\right)^2},
\label{eqks1}
\eeq
with ${\bar{m}}^n_{s}(k_\perp)=\sqrt{{k^n_\perp}^2+{m^*}^2}-s \mu_N \kappa_n B$, up to the maximum value, and integrate ${\tilde g(k^n_z(k_\perp), s)}$  over the disk $\cal{ S_{\perp}}$ in the $XY$ plane. The average of the function $\tilde g$ finally reads as,
\beq
<{\tilde g}>_s=\frac{\int_{\cal{ S_{\perp}} }{\tilde g} dS}{\int_{\cal{ S_{\perp}} }dS}
=\frac{\int_{0}^{2 \pi}\int_{0}^{\sqrt{(E^n_F+s\kappa_n \mu_N B)^2-{m^*}^2)}} {\tilde g(k^n_z(k_\perp),s)} k_\perp d k_\perp d \phi}
{\int_{0}^{2 \pi} \int_{0}^{\sqrt{(E^n_F+s\kappa_n \mu_N B)^2-{m^*}^2)}} k_\perp d k_\perp d \phi}
\label{avfng}
\eeq 

\section{Results}
\label{results}

We have used the thermodynamic conditions arising from the selfconsistent solution of the RMF set of equations Eqs.(\ref{mes1})-(\ref{mes3v}) solving beta equilibrium in a charge neutral nuclear homogeneous system under the influence of a strong quantizing magnetic field. 
\begin{figure}[hbtp]
\begin{center}
\includegraphics [angle=0,width=0.6\linewidth] {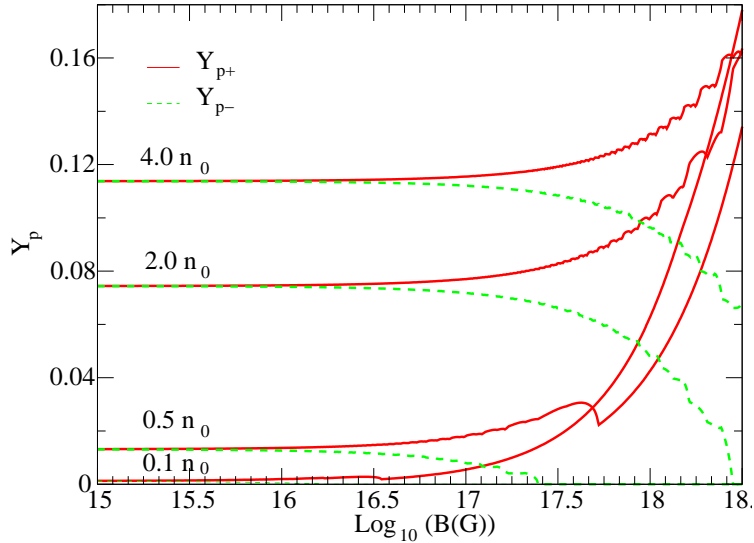}
\caption{(color online) Proton fraction for the up (red solid line) and down (green dashed line) spin  as a function of the logarithm (base 10) of the magnetic field strength at baryonic densities $n_B/n_0=0.1,0.5,2,4$.}
\label{Fig3}
\end{center}
\end{figure}
 In Fig.{\ref{Fig3}} we show proton population fractions for particles with magnetic moments polarized parallell, $(Y^{+}_p)$ (solid line),  and antiparallel, $(Y^{-}_p)$ (dashed line),  to the magnetic field, $B$, for different baryonic densities $n_B/n_0=0.1,0.5,2,4$ as a function of the logarithm (base 10) of the magnetic field strength. Due to the tiny value of the baryon magnetic moment, magnetic field strengths larger than Log B(G)$>16$  are needed so that there is a rapid increase (decrease) of the up  (down) proton fraction. For a given magnetic field strength, the differences between fractions of protons
polarized parallell and antiparallel to the magnetic field direction are larger for the smaller densities. This behavior is a consequence of the  energy balance of the interaction of
the proton  orbit magnetic momentum and  AMM  with the  magnetic field. Due to the  positive charge and the fact that $\kappa_p>0$, a quasiparticle energy is lowered by aligning spins with B, and the opposite for
spin down. Let us remind  that for densities below saturation density, $n_B=n_0=0.145$ fm$^{-3}$, the spatially non-homogeneous systems are energetically prefered over the uniform ones with the onset of {\it Pasta phases} \cite{horowitz2}, however this density can be of interest to the study of low density neutron  rich gas among the expected pasta structures.
\begin{figure}[hbtp]
\begin{center}
\includegraphics [angle=0,width=0.6\linewidth] {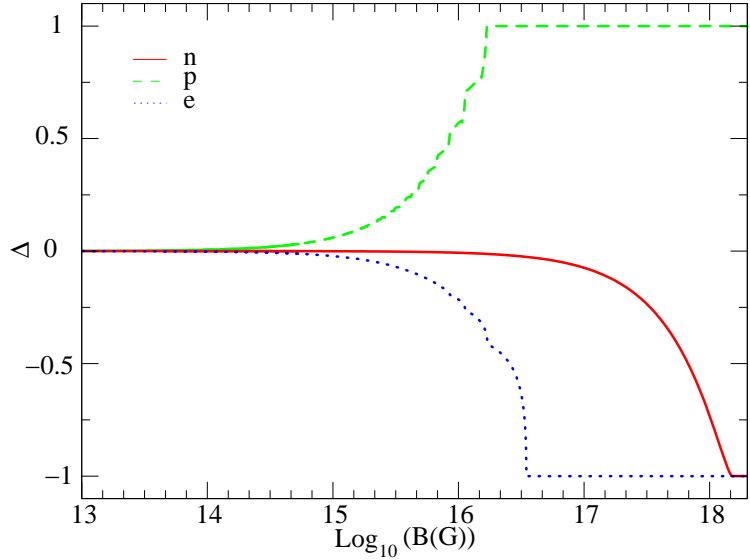}
\caption{(color online) Proton (green dashed line), neutron (red full
line) and electron (blue
dotted line) relative polarization as a function of  logarithm (base
10) of the magnetic field strength (in Gauss) for $n_B/n_0=0.1$.}
\label{Fig4}
\end{center}
\end{figure}

In Fig.{\ref{Fig4}} we show relative polarization,  $\Delta_i=(n^+_{i}- n^-_{i})/(n^+_{i}+ n^-_{i})$, where $n_i^{\pm}$ are the vector number densities of $i$-particle population, for neutrons (solid line) , protons (dashed line) and electrons (dotted line) as a function of the logarithm (base 10) of the magnetic field strength (in Gauss) at a density $n_B=0.0145$ $fm^{-3}$.  At 
this low density and for fields Log (B(G)) $>16$ there is a complete alignment of the
proton sector where they will all be in the $n=0$ Landau level, while the opposite charge of
electrons force them to be in the antialigned state with $n=0$ for a slightly smaller
field. Instead, the antipolarization of neutrons occurs due to the different interaction ($\kappa_n <0$) of the neutron AMM  with the magnetic field.  In this sense, this low density case could somewhat illustrate the properties of the homogeneous neutron gas and it remains to be seen if the interaction of the B field and other effects \cite{angPRL} could prevent the formation of proposed exotic superfluid states in the interior of NS.
 
\begin{figure}[hbtp]
\begin{center}
\includegraphics [angle=0,width=0.6\linewidth]  {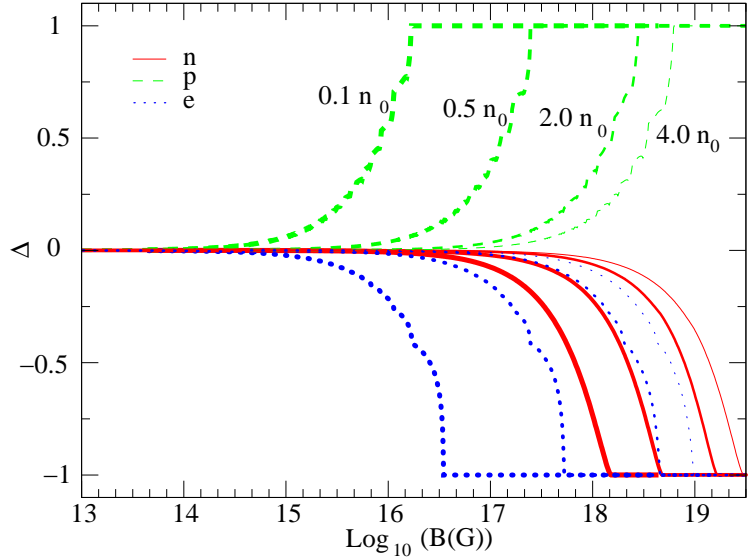}
\caption{(color online) Proton (dashed green), neutron (solid red) and electron
(dotted blue) relative polarization as a function of  logarithm (base 10) of
the magnetic field strength (in Gauss) for $n_B/n_0=0.1, 0.5, 2, 4$.}
\label{Fig5}
\end{center}
\end{figure}
In Fig.{\ref{Fig5}} we show relative polarization, $\Delta$,  for
neutrons (red  solid line) , protons (green dashed line) and electrons (blue dotted line) as a function of the logarithm (base 10) of the magnetic field
strength (in Gauss) for $n_B/n_0=0.1, 0.5,  2,  4$. Increasing densities are depicted with decreasing line width. Hadrons and leptons show different behavior according to the associated sign and value of their magnetic
moment. For low density at $n_B/n_0=0.1$ protons show a total polarization for magnetic
fields larger than $B \approx 10^{16}$ G while for electrons the polarization is slightly smaller  and  with opposite relative sign because of its negative electrical charge. From this figure it is seen that as density grows the magnetic field strength needed to polarize a given population fraction is bigger, due to competing effects in Fermi energies. The different behavior between protons
and neutrons is due to the orbital magnetic moment contributing only for protons. In general, it is quite difficult to polarize  neutrons, and
$\Delta_n$ remains always below $\Delta_p$ (in absolute value). 

\begin{figure}[ht]
\begin{center}
\makebox[0pt]{\includegraphics [angle=0,scale=.6] {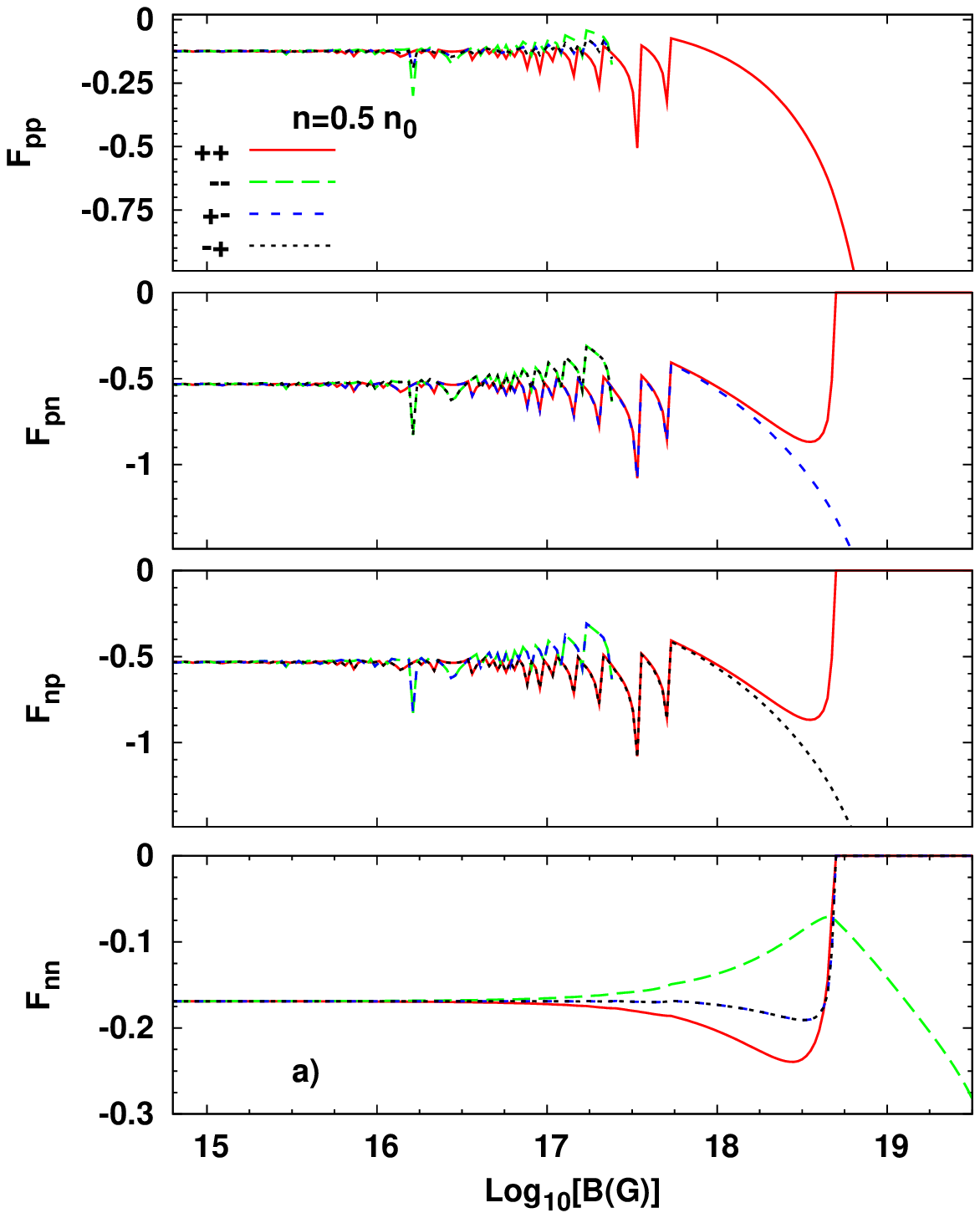}
\kern 5mm \includegraphics [angle=0,scale=.6] {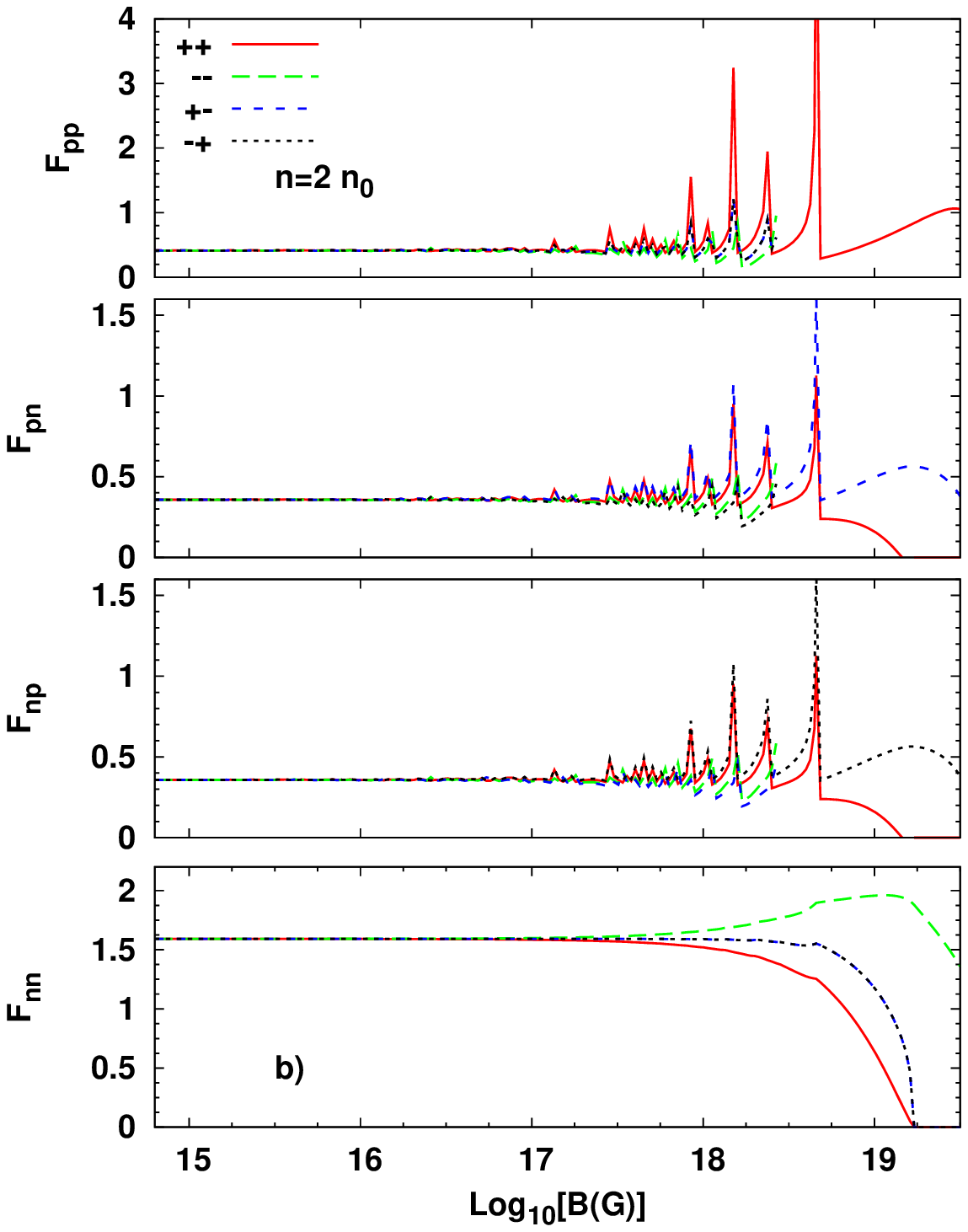}}\\[0mm]
\caption{\small \label{Fig6}
(color online) Averaged and normalized Landau parameters $F^{s s'}_{ij}$ as functions of the logarithm of the magnetic field strength for $s,s'=\pm 1$ at densities $n_B/n_0=0.5$ (left pannel) and $n_B/n_0=2$ (right pannel).}
\end{center}
\end{figure}

In Fig.{\ref{Fig6} we show the averaged and normalized Landau parameters $F^{ss'}_{ij}$ given by Eq.(\ref{def_f}) and Eq.(\ref{avfng}) for the different combinations of isospin  $\{i,j\}=\{n,p\}$ as functions of the logarithm (base 10) of the magnetic field strength (in Gauss) for: a) $n_B/n_0=0.5$  (left pannel) and b) $n_B/n_0=2$ (right pannel). $F_{pp}$, $F_{pn}$, $F_{np}$ and $F_{nn}$ are depicted from top to bottom in each pannel. We can see the different polarization components, $(++$),($+-$),($-+$), ($--$), in solid, long dashed, short dashed and dotted lines respectively. The peaks appearing in the dimensionless coefficients involving protons, that is, $(F_{pp},F_{pn}, F_{np})$, are due to the fact the level density depends on the Landau level, $n$, and spin projection, $s$. Instead, for neutrons the behavior is smooth since Fermi surfaces are ellipsoids.
We can see that the parameters $F^{ss'}_{ij}$ for the low density case on the left pannel are negative, signaling  attractive interaction, while for the high density case  they are always positive, or repulsive,  for the magnetic field strengths considered in this work.  Results for B field strengths with Log$_{10}$(B(G)) $>18$ should be taken with care and considered as an extrapolation. At low B the hopping behavior is smooth due to the large number of Landau levels populated and as B grows the number of levels decreases and the separation between them increases (see Fig. \ref{Fig1}).  For sufficiently large B, $k^p_{F,\nu,s=-1}=0$, and beyond that strengh the density of states, $N^-_p$ and  all components in the $F_{pp}$ except for the $(++)$ are not defined. The same happens for $F_{pn}, F_{np}$, since they involve the proton spin down component. For these components there is a symmetry by replacing simultaneously spin and isospin indexes, in this way, for instance,  the $F^{+-}_{pn}$ and $F^{-+}_{np}$ are equal. For strengths of $B \approx 4 \times 10^{18}$ G all the neutrons are polarized down and the only non vanishing components are those involving this fraction.  

At low densities and for components with mixed spin the quasiparticle states are more bound than same spin components. Protons tend to polarize up, as B grows, filling the low Landau levels and the qp  interaction becomes more attractive, however for neutrons this behavior reverses since, the larger B, the smaller the neutron fraction populates the system and the more repulsive the medium reacts to the formation of a qp state. Then, this causes quasineutron excitations to be less bound on an absolute scale. In the high density case (right pannel), the spikes signal the hopping of the Landau levels, and the clear more repulsive interaction than  shown on the low density case (left pannel) where the magnetization is stronger. The mixed states involving protons are less repulsive than the $F_{nn}$ in the mostly neutron populated system. 
\begin{figure}[ht]
\begin{center}
\makebox[0pt]{\includegraphics [angle=0,scale=.6] {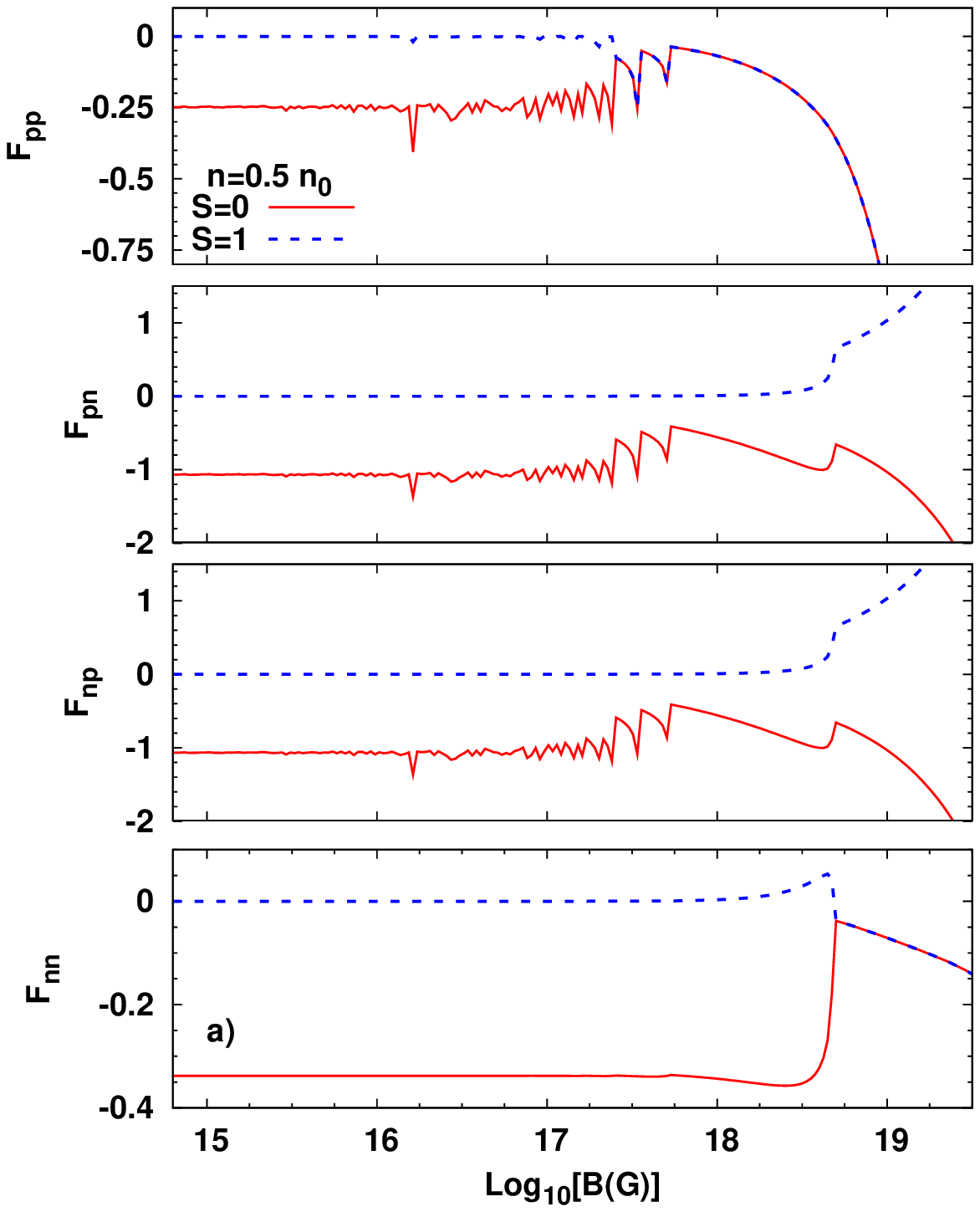}
\kern 5mm \includegraphics [angle=0,scale=.6] {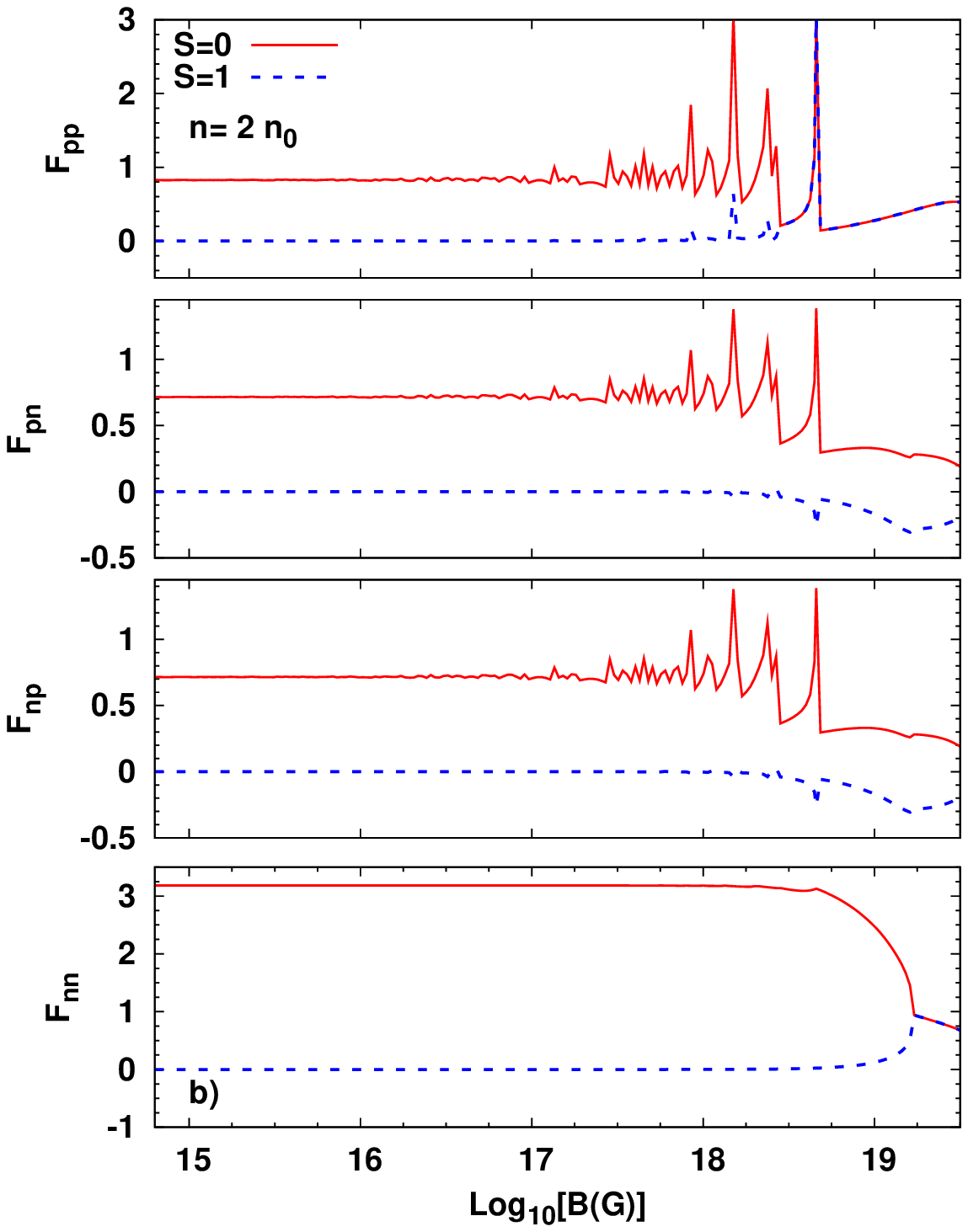}}\\[0mm]
\caption{\small \label{Fig7}
(color online)Averaged and normalized Landau parameters $F^{S}_{ij}$ as functions of the logarithm of the magnetic field strength for $S=0, 1$ for densities $n_B/n_0=0.5$ (left pannel) and $n_B/n_0=2$ (right pannel).}
\end{center}
\end{figure}

In Fig.{\ref{Fig7} we show the $F^{S}_{ij}$ normalized Landau coefficients  according to the definitions in Eq.(\ref{fs0s1}) with a definite value of total spin, $S$, for all combinations of isospin $\{i,j\}=\{n,p\}$ as a function of the logarithm (base 10) of the magnetic field strength (in Gauss) for  a) $n_B/n_0=0.5$ (left pannel) and b) $n_B/n_0=2$ (right pannel). $F_{pp}$, $F_{pn}$, $F_{np}$ and $F_{nn}$ are depicted from top to bottom in each pannel. The singlet (solid line)  and triplet (dashed line) state parameters present a different behavior due to their construction. As seen in the left pannel for the low density case, due to cancellations of negative components $(++,+-,-+,--)$, the  triplet states have small binding. Notice that for the $F_{pp}$ the well-defined values are those where $k^p_{F,\nu,s}$ is real ($\nu \le \nu_{max}$). For singlet states the energy is lowered with respect to triplet states, for different isospin components($pn, np$). At higher densities, $n_B/n_0=2$ (right pannel), the different components $F^{ss'}_{ij}$ are all positive, and, therefore, the singlet and triplet binding behave in a similar way  to the low density case, however, in absolute value the singlet case will have repulsive character.
This analysis of Landau parameters could be of potential interest to obtain transport coefficients from the components of their expansion in cylindrical harmonics, due to  the axial symmetry introduced by the existence of a strong quantizing magnetic field. From this expansion information of low temperature nuclear system observables \cite{chamel08} such as, for example,  effective mass, spin diffusivity, viscosity, etc could be accessed. Additional contributions from exchange terms to the calculated parameters should also be considered.

\section{Summary}
\label{summary}

In the present paper we have derived for the first time the Landau parameters for charge neutral homogeneous asymmetric nuclear matter under very strong magnetic fields. In particular, we have considered cold stellar matter in $\beta$-equilibrium in the presence of fields with strengths Log$_{10}$ (B(G))$<18$ as allowed by the scalar virial theorem.  We have used a RMF model to describe the nuclear system so that our
calculation may be extended to high densities. However, no exotic matter such as hyperons or
kaon condensates has been considered. In our work, we have only considered matter densities from subsaturation density, $n_B/n_0=0.1$, to densities before a possible quark transition.  Present results at subsaturation densities should be taken
with care,  as a first approach to a realistic description of the effect of strong magnetic fields on the cold stellar matter. Also we have not considered the existence
of superfluid neutrons at subsaturation densities.

We find that the nuclear system needs at least magnetic field strengths larger than $B \approx 10^{16}$ G to show some noticeable magnetization in excitation spectrum. Protons tend to polarize aligning spins with B field direction, while neutrons and electrons do the opposite. Landau coefficients $F^{s s'}_{ij}$ show the interaction energy of qp excitations with isospins $i$ and $j$ and spins $s$ and $s'$, respectively, and are calculated from the energy density ${\cal E}$ of the system in charge neutrality as a two body operator obtained from its second derivative with respect to the occupation numbers of the qp excitation. 
In the excitations involving quasiprotons the discrete Landau levels play an important role quantizing the momentum states in the direction perpendicular to the field. In this way there is a discrete feature associated in the Landau coefficients. For small B the number of allowed levels is large and is hardly visible on the plots, however for higher B the number of levels decreases and they are more separated. For neutrons instead, the lack of quantized Landau levels implies a smooth behavior of the parameters as functions of magnetic field strength.

At low densities the effect of the magnetic field is small on the individual polarization since the polarization induced is weak (for fields with Log$_{10} (B(G))<17$) and gives an attractive overall effect with respect to unmagnetized systems. The dominant component drives the tendency of the parameter, in this way the proton sector is more bound and so are the $pp,pn,np$ components, while for neutrons they are less bound since the number of neutrons decreases when B increases. In this case, for situations of complete polarization the Landau parameters get largely negative. In the high density cases the interaction is increased in the way to unbind the qp excitations.The medium effects are such that the interaction is repulsive and the magnetic field tends to decrease this effect in the less populated fractions while increasing in the more populated ones.

For the singlet qp configuration the medium effects play a role of
binding at low densities while at high densities the opposite occurs.
The B field tends to reverse this tendency (if not too strong) and can
be considered a competing effect. As for the triplet the qp spin
configurations  the in-medium effects play no role, while the B field
brings extra binding in the pp or nn channel at low densities, while
the opposite occurs at high densities. The effect of the B field is
reversed in the pn or np channels, binding (unbinding) is present at
high (low) densities.

The calculation we have performed could have potential interest to size the relevance of the inclusion of a magnetic field in astrophysical scenarios with realistic description  of nuclear systems in beta equilibrium by relativistic mean field theoretical models. We have obtained singlet and triplet interaction energies of qp states and further work could relate these, in a consistent fashion,  to relativistic and (possibly) superfluid hadronic components in this type of systems.
\newpage
\section{Appendix}
\label{appendix} 

In this appendix  we provide the explicit form of the quasiparticle interaction matrix elements depending on spin and isospin and the mass and current derivatives appearing in the calculation of the Landau parameters in section \ref{landaucoef}.

\subsection{Interaction matrix elements}
The matrix elements appearing in $f_L$ in Eq.(\ref{matrixf}) in section \ref{landaucoef} read as:

i) for quasiproton-quasiproton  interaction,
\bea
f^{pp}_{ilss'\nu\nu'}&=&\frac{\partial \epsilon^{p}_{i\nu, s}}{\partial n^p_{ls'\nu'}} \nonumber\\
&=& \frac{\bar m^p_{\nu s}}{\epsilon^p_{i\nu s}} \frac{\partial \bar m^p_{\nu s}}{\partial
  n^p_{ls'\nu'}}- \frac{{k}_{zi}}{\epsilon^p_{i\nu s}}  \frac{\partial {\cal V}^p_{z}}{\partial
  n^p_{ls'\nu'}} + A_+ \nonumber\\
&=& \frac{\bar m^p_{\nu s}}{\epsilon^p_{i\nu s}}  \frac{m^*}{\bar m^p_{\nu s}+s\kappa_p \mu_N B}\frac{\partial m^*}{\partial
  n^p_{ls'\nu'}}- \frac{{k}^p_{zi}}{\epsilon^p_{i\nu s}} \left[A_+\frac{\partial
j^p_{z}}{\partial  n^p_{ls'\nu'}} +A_-\frac{\partial j^n_{z}}{\partial  n^p_{ls'\nu'}} \right]
+ A_+ \nonumber\\
&=& -\frac{g_\sigma^2}{{m'_\sigma}^2}
\frac{\bar m^p_{\nu s}}{\epsilon^p_{i\nu s}}  
\frac{m^*}{\tilde \epsilon^p_{\nu s}}
\frac{\bar m^p_{\nu' s'}}{\epsilon^p_{l\nu' s'}}  
\frac{m^*}{\tilde \epsilon^p_{\nu' s'}}
\frac{1}{1+\frac{g_\sigma^2}{{m'_\sigma}^2} \left(I_1+I_2\right)}
- \frac{{k}^p_{zi}}{\epsilon^p_{i\nu s}}
  \frac{{k}^p_{zl}}{\epsilon^p_{l\nu' s'}}
\frac{A_++\left( A_+^2-A_-^2\right)b_n}{D}\nonumber\\
&+&A_+
 \label{fpp}
\eea
ii) for quasineutron-quasineutron  interaction,
\bea
f^{nn}_{ilss'}&=&\frac{\partial \epsilon^{n}_{i s}}{\partial n^n_{ls'}} \nonumber\\
&=&   A_+ 
- \frac{{k}^n_{zi}}{\epsilon^n_{i s}}  \frac{\partial {\cal V}^n_{z}}{\partial n^n_{ls'}}
+\frac{\epsilon^n_{\perp i}-s\kappa_n \mu_N B}{\epsilon^n_{ is}{\epsilon^n_{\perp i}} }\left[{m^*}\frac{\partial m^*}{\partial
  n^n_{ls'}}-{\mathbf {k}^n_{i\perp s}}
\frac{\partial {\mathbf {\cal V}}^n_{\perp } }{\partial n^n_{ls'}}\right]\nonumber\\
&=&   A_+ 
- \frac{{k}^n_{zi}}{\epsilon^n_{i s}}  \left[A_-\frac{\partial j^p_{z}}{\partial  n^n_{ls'}}
+A_+\frac{\partial j^n_{z}}{\partial  n^n_{ls'}} \right] +\frac{\epsilon^n_{\perp i}-s\kappa_n
\mu_N B}{\epsilon^n_{ is}{\epsilon^n_{\perp i}}}\left[{m^*}
\frac{\partial m^*}{\partial
  n^n_{ls'}}-\boldsymbol{k}^n_{i\perp}\left(A_+\frac{\partial
\boldsymbol{j}^n_{\perp}}{\partial  n^n_{ls'}} \right)\right]\nonumber\\
&=&   A_+ 
- \frac{{k}^n_{zi}}{\epsilon^n_{i s}}   \frac{{k}^n_{zl}}{\epsilon^n_{l s'}}
\frac{A_++\left( A_+^2-A_-^2\right)\,b_p} {D} \nonumber\\
&-&\frac{g_\sigma^2}{{m'_\sigma}^2}
\frac{\left(\epsilon^n_{\perp i}-s\kappa_n \mu_N B\right)\, m^*}{\epsilon^n_{ is}{\epsilon^n_{\perp
i}}}
\frac{\left(\epsilon^n_{\perp l}-s'\kappa_n \mu_N B\right)\, m^*}{\epsilon^n_{ ls'}{\epsilon^n_{\perp
l}}} \frac{1}{1+\frac{g_\sigma^2}{{m'_\sigma}^2} \left(I_1+I_2\right)}\nonumber\\
&-&\frac{\left(\epsilon^n_{\perp i}-s\kappa_n \mu_N B\right)}{\epsilon^n_{ is}{\epsilon^n_{\perp
i}}}
\frac{\left(\epsilon^n_{\perp l}-s'\kappa_n \mu_N B\right)}{\epsilon^n_{ ls'}{\epsilon^n_{\perp
l}}} \boldsymbol{k}^n_{i\perp}\cdot  \boldsymbol{k}^n_{l\perp} \frac{A_+}{1+A_+ I_3}
 \eea
iii) for quasiproton-quasineutron interaction,
\bea
f^{pn}_{ilss'\nu}&=&\frac{\partial \epsilon^{p}_{i s\nu}}{\partial n^n_{ls'}} \nonumber\\
&=&   A_- 
+ \frac{m^*}{\bar m^p_{\nu s}+s\kappa_p \mu_N B} \frac{\bar m^p_{\nu s}}{\epsilon^p_{i\nu s}} 
\frac{\partial \bar m^*}{\partial n^n_{ls'}}- \frac{{k}_{zi}}{\epsilon^p_{i\nu s}} \left[A_+\frac{\partial j^p_{z}}{\partial  n^n_{ls'}} +A_-\frac{\partial j^n_{z}}{ \partial n^n_{ls'}} \right] \nonumber\\
&=&   A_- 
-\frac{g_\sigma^2}{{m'_\sigma}^2} \frac{m^{*2}\, \bar m^p_{\nu s}}{\epsilon^p_{i\nu s}\tilde \epsilon^p_{\nu s}} 
\frac{\left(\epsilon^n_{\perp l}-s'\kappa_n \mu_N B\right)\, m^*}{\epsilon^n_{ ls'}{\epsilon^n_{\perp
l}}}
 \frac{1}{1+\frac{g_\sigma^2}{{m'_\sigma}^2} \left(I_1+I_2\right)}
+ \frac{{k}_{zi}}{\epsilon^p_{i\nu s}}  \frac{{k}_{z}}{\epsilon^n_{l s'}} \frac{A_-}{D}
 \eea
iii) for quasineutron-quasiproton interaction,
\bea
f^{np}_{ilss'\nu'}&=&\frac{\partial \epsilon^{n}_{i s}}{\partial n^p_{ls'\nu'}} \nonumber\\
&=&   A_- 
- \frac{{k}_{zi}}{\epsilon^n_{i s}}  
\left[A_-\frac{\partial j^p_{z}}{\partial  n^p_{ls'\nu'}} +A_+\frac{\partial j^n_{z}}{
 \partial   n^p_{ls'\nu'}} \right] 
+ \frac{\epsilon^n_{\perp i}-s\kappa_n \mu_N B}{\epsilon^n_{ is}{\epsilon^n_{\perp i}}}
\left[{m^*}\frac{\partial m^*}{\partial  n^p_{ls'\nu'}}
-\boldsymbol{k}_{i\perp } 
\left(A_+\frac{\partial \boldsymbol{j}^n_{\perp}}{  \partial  n^p_{ls'\nu'}} \right)
\right]\nonumber\\
&=&   A_-
 -\frac{g_\sigma^2}{{m'_\sigma}^2}
 \frac{\epsilon^n_{\perp i}-s\kappa_n \mu_N B}{\epsilon^n_{ is}{\epsilon^n_{\perp i}}}
 \frac{m^{*2}\, \bar m^p_{\nu' s'}}{\epsilon^p_{l\nu' s'}\tilde \epsilon^p_{\nu' s'}} 
 \frac{1}{1+\frac{g_\sigma^2}{{m'_\sigma}^2} \left(I_1+I_2\right)}+ 
\frac{{k}_{zi}}{\epsilon^n_{i s}}
\frac{{k}_{zl}}{\epsilon^p_{i s' \nu'}} \frac{A_-}{D}.
 \label{fnp}
 \eea
 The indexes in labels in expressions Eqs.(\ref{fpp})-(\ref{fnp}) have been accordingly written  with six numbers $(i,\nu, s), (l, \nu',s')$ for quasiproton interaction, and  four numbers  $(i, s), (l,s')$ for quasineutron interaction. For isospin mixing terms (quasiproton (quasineutron)-quasineutron (quasiproton)) there are 5 numbers $(i, \nu, s), (l, s')$ in the label. We have used auxiliar definitions in the combination of coupling constants,
\beq
A_\pm=\frac{g_\omega^2}{m_\omega^{'2}}\pm\frac{g_\rho^2}{4\, m_\rho^2},
\eeq
 and the proton effective energy
\beq
\epsilon^p_{i\nu s}=\sqrt{k^{2}_{zi}+  (\bar m^p_{\nu s})^{2}},
\eeq
with an auxiliar mass $\bar m^p_{\nu s}=\sqrt{{m^*}^2 + 2 q_p \nu B}- s\kappa_p \mu_N B=\tilde \epsilon^p_{\nu s}- s\kappa_p \mu_N B$.

For neutrons we define an auxiliar energy (which can be interpreted as a neutron momentum dependent mass):
\beq
\epsilon^n_{\perp i}=\sqrt{m^{* 2}_{n}+k^{2}_{\perp i}},
\eeq
and the effective energy for neutrons:
\beq
\epsilon^n_{i s}=\sqrt{k^{2}_{zi}+\left(\epsilon^n_{\perp i}-s\kappa_{n} \mu_{N} B
\right)^{2}}.
\eeq
We also define,
  $$D={\left(1+A_+\, b_n\right)\left(1+A_+\, b_p\right)- A_-^2 \, b_p\, b_n},$$
where the coefficients $b_p$ and $b_n$ are:
\beq
b_p=\sum_j  \frac{\displaystyle n^p_{js^{\prime\prime}\nu^{\prime\prime}}}
{\displaystyle \epsilon^p_{js^{\prime\prime}\nu^{\prime\prime}}}
\left[1- \left(\frac{\displaystyle k_{jz}}
{\displaystyle \epsilon^p_{js^{\prime\prime}\nu^{\prime\prime}}}\right)^2
\right]= \frac{q_p \, B}{2\pi^2}\sum_{s,\nu}\frac{k^p_{F\nu s}}{E^p_{F}},
\eeq
\bea
b_n&=&\sum_j  \frac{\displaystyle n^n_{js^{\prime\prime}}}
{\displaystyle \epsilon^n_{js^{\prime\prime}}}
\left[1- \left(\frac{\displaystyle k_{jz}}
{\displaystyle \epsilon^n_{js^{\prime\prime}}}\right)^2
\right]\nonumber\\
&=& \frac{1}{2\pi^2}\sum_s\left\{\frac{(k^n_{F s})^3}{3 E^n_{F}}-\frac{s \kappa_n \mu_N B}{
2 E^n_{F}}\left[\bar m k^n_{F s} - E^{n 2}_{F}
\left[\frac{\pi}{2}-\arcsin\left(\frac{\bar m_n}{ E^n_{F}}\right)\right]\right]
\right\},
\eea
and  $k^n_{Fs}=\sqrt{{E^n_{F}}^2-{\bar m_n}^2}$. The explicit form for the integrals $I_1,\, I_2$ and $I_3$ is given by

\bea
I_1&=&\sum_{j\nu^{\prime\prime} s^{\prime\prime}}
n^p_{j\nu^{\prime\prime} s^{\prime\prime}} 
\frac{\displaystyle \partial^2 \epsilon^p_{j\nu^{\prime\prime} s^{\prime\prime}}}{\displaystyle\partial {m^*}^2}\nonumber \\
&=& \frac{q_p \, B}{2\pi^2} \sum_{\nu, s}
 \int_{0}^{k^p_{F,s}} dx \frac{1}{\sqrt{k^2_z+{\bar m_p}^2}}
 \left(\frac{m^{2*}(s\kappa_p \mu_N B)}
{ ({\tilde\epsilon}^p_{\nu s})^3}+\frac{{\bar m_p}}{{\tilde\epsilon}^p_{\nu s}} -\frac{{\bar
m_p} m^{*2}}{\left({\tilde\epsilon}^p_{\nu s}\right)^2}
 \frac{1}{\left(\sqrt{k^2_z+{\bar m_p}^2}\right)^2}\right) \nonumber\\
&=& \frac{q_p \, B}{2\pi^2} \sum_{\nu, s}
\left[\frac{\bar m^p_{\nu s}2 q_p \nu B + {m^*}^2 \tilde\epsilon^p_{\nu s} }
{\left(\tilde\epsilon^p_{\nu s}\right)^3}
\ln \left(\frac{E^p_{F}+k^p_{F\nu s}}{\bar m^p _{\nu s}}\right)-\frac{{m^*}^2 {k^p_{F\nu s}}}{E^p_{F} \left({\tilde\epsilon}^p_{\nu s}\right)^2}
\label{i1}
\right]
\eea

\bea
I_2&=& \sum_{j s^{\prime\prime}}
n^n_{j s^{\prime\prime}} 
\frac{\displaystyle \partial^2 \epsilon^n_{j s^{\prime\prime}}}
{\displaystyle\partial {m^*}^2}\\
&=& \sum_s\int_{\bar m_n}^{E^n_{F}} dx ( x+s \kappa_n \mu_N B) \left\{\frac{  {m^*}^2 \, s \kappa_n \mu_N B+ x\left(x+s \kappa_n \mu_N B\right)^2  }{\left(x+s \kappa_n \mu_N B\right)^3}
\ln\left[\frac{\sqrt{(E^n_{F})^2-x^2}+E^n_{F}}{x}\right]\right.\nonumber \\
&&\left.- \frac{{m^*}^2\sqrt{(E^n_{F})^2-x^2}}{(x+s\kappa_n \mu_N B)^2 E^n_{F} } \right\}\nonumber\\
&=&{\frac{1}{2\pi^2}\sum_s\int_{\bar m_n}^{E^n_F} dx\left\{
\left[\frac{{m^*}^2\, s \kappa_n \mu_N B}{(x+s\kappa_n \mu_N B)^2}+x \right]
\ln\left(\frac{E^n_F+\sqrt{(E^n_F)^2-x^2}}{x}\right)- \frac{{m^*}^2}{E^n_F}
\frac{\sqrt{(E^n_F)^2-x^2}}{{x+s\kappa_n \mu_N B}}
\right\}}.\nonumber
\label{i2}
\eea

The integral $I_2$ can be written out into several pieces $I_2=I_2^a+I_2^b+I_2^c$.
\bea
I_2^a&=& \frac{1}{2\pi^2}\sum_s\int_{\bar m_n}^{E^n_F} dx
\frac{{m^*}^2\, s \kappa_n \mu_N B}{(x+s\kappa_n \mu_N B)^2}
\ln\left(\frac{E^n_F+\sqrt{(E^n_F)^2-x^2}}{x}\right)\\
I_2^a&=& \sum_{s} -\frac{(s \kappa_n \mu_N B)m^{*2}}{2 \pi^2}
\left\{
\frac{E^n_{Fs}}{(s \kappa_n \mu_N B)\alpha}
\left[
 \ln \left(\frac{(E^n_{Fs})^2+s \kappa_n \mu_N B E^n_{Fs}}
{\alpha k_{Fs}+(E^n_{Fs})^2+s \kappa_n \mu_N B{\bar m}_n}\right) \right. \right. \nonumber \\
&-& \left. \ln \left(\frac{E^n_{F}+s \kappa_n \mu_N B }{{\bar m}+s \kappa_n \mu_N B}\right)
\right] -
\frac{1}{(s \kappa_n \mu_N B)}
\left[
\ln\left(\frac{\bar E^n_F}{E^n_{Fs}+k_{Fs}}\right)
-\ln\left(\frac{E^n_{Fs}}{{\bar m}}\right)
\right]
\nonumber \\
&-& \left.
\frac{1}{(s \kappa_n \mu_N B+{\bar m})}
\ln\left(\frac{k_{Fs}+E^n_{Fs}}{{\bar m}}\right)
\right\},
\label{I2acp}
\eea

\beq
I_2^b= \frac{1}{2\pi^2}\sum_s\int_{\bar m_n}^{E^n_F} dx  x\,\ln \left( \frac{ E^n_F +\sqrt{(E^n_F)^2-x^2}}{x}\right),
\eeq

\beq
I_2^b
=\frac{-1}{4 \pi^2}\sum_{s} \left[ {\bar m}^2 \ln \left(\frac{k^n_{Fs}+E^n_{F}}{{\bar m}}\right)-
k^n_{Fs} E^n_{F} \right] .  \qquad
\label{I2b}
\eeq

\bea
I_2^c&=& \frac{-m^{*2}}{2 \pi^2 }\sum_{s}\int_{\bar m_n}^{E^n_F} dx \,\frac{\sqrt{(E^n_F)^2-x^2}}{ E^n_F (x+s\kappa_n \mu_N B )},
\eea
\bea
I_2^c&=& \frac{-m^{*2}}{2 \pi^2 E^n_{F}}\sum_{s} 
\left\{
 -\alpha \ln \left(\frac{(E^n_{F})^2+s \kappa_n \mu_N B E^n_{F}}{\alpha k^n_{F}+(E^n_{F})^2+s \kappa_n \mu_N B {\bar m}}\right)
+\alpha \ln \left(\frac{E^n_{F}+s \kappa_n \mu_N B }{{\bar m}+s \kappa_n \mu_N B}\right)\right.
\nonumber\\
&+&\left.
s \kappa_n \mu_N B \left[\frac{\pi}{2}-arctan \left(\frac{\bar m}{k^n_{Fs}}\right) \right]-k^n_{Fs}\right\},
\label{I2c}
\eea
with $\alpha=\sqrt{(E^n_{F})^2-(s \kappa_n \mu_N B)^2}$. The integral $I_3$ can be written as,
\bea
I_3&=&\sum_{j s^{\prime\prime}}n^n_{j s^{\prime\prime}}
\left[
\frac{\displaystyle \epsilon^n_{j\perp}-s^{\prime\prime} \kappa_n \mu_N B}
{\displaystyle \epsilon^n_{j s^{\prime\prime} }\epsilon^n_{j\perp}}
\left( 1- \frac{\displaystyle \epsilon^n_{j\perp}-s^{\prime\prime} \kappa_n\mu_N B}
{\displaystyle {2\epsilon^n_{j s^{\prime\prime}}}^2 {\epsilon^n_{j\perp}}} k^2_{j\perp}
\right)
+\frac{\displaystyle s^{\prime\prime} \kappa_n \mu_N B}
{\displaystyle 2 \epsilon^n_{j s^{\prime\prime} }{\epsilon^n_{j\perp}}^3} k^2_{j\perp}
\right].
\eea
We write the integral as several pieces $I_3=I_3^a+I_3^b+I_3^c$. For $I_3^a$ we have
 $I_3^a=I_2^b$.
For $I_3^b$,
\bea
I_3^b&=& \frac{-1}{4 \pi^2 E^n_{F}}\sum_s\int_{\bar m_n}^{E^n_{F}} dx \left\{
\frac{\left[\left(x+s \kappa_n \mu_N B\right)^2-{m^*}^2\right]}{\left(x+s \kappa_n \mu_N B\right)} \sqrt{(E^n_{F})^2-x^2}
\right\},
\eea
\bea
 I_3^b &=&\frac{-1}{4\pi^2 E^n_F}\sum_s \left\{
 \frac{1}{3}(k_{F s})^3-
\frac{s \kappa_n \mu_N B}{2} {\bar m} k^n_{F s}+(E^n_{F})^2\frac{s \kappa_n \mu_N B}{2}
\left[\frac{\pi}{2}-\mbox{arctan}\left(\frac{\bar m_n}{k_{F s}}\right)\right]\right\} 
\nonumber \\
&+&\frac{m^{*2}}{4 \pi^2 E^n_{F}} \sum_s \left\{ -\alpha 
\left[\ln \left(\frac{(E^n_{F})^2+s\kappa_n \mu_N B E^n_{F}}{\alpha k_{Fs}+(E^n_{F})^2+s \kappa_n \mu_N B {\bar m_n}}\right)
- \ln \left(\frac{E^n_{F}+s \kappa_n \mu_N  B }{{\bar m_n}+s \kappa_n  \mu_N  B}\right)\right]\right. \nonumber \\
&+& \left.
s \kappa_n \mu_N  B \left[\frac{\pi}{2}-\mbox{arctan} \left(\frac{\bar m}{k_{Fs}}\right)\right]-k_{Fs} \right\}.
\label{i3b}
\eea
For $I_3^c$,
\bea
I_3^c&=& \frac{s\kappa_n \mu_N B}{4 \pi^2}\sum_s\int_{\bar m_n}^{E^n_{F}} dx \left\{
\frac{\left[\left(x+s \kappa_n \mu_N B\right)^2-{m^*}^2\right]}{\left(x+s \kappa_n \mu_N B\right)^2}
\ln\left(\frac{\sqrt{(E^n_{F})^2-x^2}+E^n_F}{x}\right)
\right\},
\eea
\bea
 I_3^c &=&\frac{s \kappa_n \mu_N B}{4\pi^2}\sum_s \left\{ 
-{\bar m}\, \ln \left(\frac{E^n_{F}+k_{Fs}}{{\bar m}}\right) +E^n_{F} \left(\frac{\pi}{2}-arctan \left(\frac{{\bar m}}{k_{Fs}}\right) \right) \right\} \nonumber \\
&-&\frac{s \kappa_n \mu_N B m^{*2}}{4\pi^2} \sum_s \left\{ \frac{-E^n_{F}}{s \kappa_n \mu_N B \alpha}
\left[\ln \left(\frac{(E^n_{F})^2+s \kappa_n  \mu_N B E^n_{Fs}}{\alpha k_{Fs}+(E^n_{F})^2+s \kappa_n \mu_N B {\bar m}}\right)
-\ln \left(\frac{E^n_{F}+s \kappa_n \mu_N B }{{\bar m}+s \kappa_n \mu_N B}\right)\right]\right\}\nonumber \\
&+&\left.
 \frac{1}{s \kappa_n \mu_N B}\left[\ln \left(\frac{E^n_{F}}{k_{Fs}+E^n_{F}}\right)-\ln \left(\frac{E^n_{F}}{{\bar m}}\right)\right]
+\frac{1}{s \kappa_n \mu_N B+{\bar m}}\ln \left(\frac{k_{Fs}+E^n_{F}}{{\bar m}}\right) \right\}.
\label{i3a}
\eea

\subsection{Effective mass derivatives}

The mass derivative appearing in eqs.(\ref{fpp}-\ref{fnp}) 
$\frac{\partial m^*}{\partial n^p_{ls'\nu'}}=\frac{\partial (m_b-g_{\sigma} \sigma)}{\partial n^p_{ls'\nu'}}$ can be obtained using the equation of motion for the scalar $\sigma$ field in Eq. (\ref{mes1}). Then we can write

\bea
\frac{\partial m^*}{\partial n^p_{ls'\nu'}}&=&
-\frac{g_\sigma^2}{{m'}^2_\sigma}\,\,\,
 \frac{\displaystyle\bar m^p_{\nu' s'} \, m^*}
{\displaystyle\epsilon^p_{l\nu's'}\, \tilde\epsilon^p_{\nu' s'}}
\frac{\displaystyle 1}
{\displaystyle 1+ \frac{g_\sigma^2}{{m'}^2_\sigma}\, 
\left(\sum_{j\nu^{\prime\prime} s^{\prime\prime}}
n^p_{j\nu^{\prime\prime} s^{\prime\prime}} 
\frac{\displaystyle \partial^2 \epsilon^p_{j\nu^{\prime\prime} s^{\prime\prime}}}{\displaystyle\partial {m^*}^2}
+ \sum_{j s^{\prime\prime}}
n^n_{j s^{\prime\prime}}
\frac{\displaystyle \partial^2 \epsilon^n_{j s^{\prime\prime}}}
{\displaystyle\partial {m^*}^2}
\right)},
\eea

\bea
\frac{\partial m^*}{\partial n^n_{ls'}}&=&
-\frac{g_\sigma^2}{{m'}^2_\sigma}\,\,\,
\frac{\left(\epsilon^n_{l\perp}-s^{\prime} \kappa_n  \mu_N  B\right)m^*}
{\displaystyle   \epsilon^n_{l s^{\prime}} \epsilon^n_{l\perp}}
\frac{\displaystyle 1}
{\displaystyle 1+ \frac{g_\sigma^2}{{m'}^2_\sigma}\, 
\left(\sum_{j\nu^{\prime\prime} s^{\prime\prime}}
n^p_{j\nu^{\prime\prime} s^{\prime\prime}} 
\frac{\displaystyle \partial^2 \epsilon^p_{j\nu^{\prime\prime} s^{\prime\prime}}}{\displaystyle\partial {m^*}^2}
+ \sum_{j s^{\prime\prime}}
n^n_{j s^{\prime\prime}}
\frac{\displaystyle \partial^2 \epsilon^n_{j s^{\prime\prime}}}
{\displaystyle\partial {m^*}^2}
\right)}.
\eea

In the limit where sums are converted into integrals it is found that,
\beq
\frac{\partial m^*}{\partial n^p_{ls'\nu'}}=
-\frac{g_\sigma^2}{{m'}^2_\sigma}\,\,
 \frac{\displaystyle\bar m^p_{\nu' s'} \, m^*}{\displaystyle
\epsilon^p_{l\nu's'}\, \tilde\epsilon^p_{\nu' s'}}
\frac{1 }{\displaystyle 1+ \frac{g_\sigma^2}{{m'}^2_\sigma}\, 
\left(I_1+I_2\right)},
\eeq

\beq
\frac{\partial m^*}{\partial n^n_{ls'}}=
-\frac{g_\sigma^2}{{m'}^2_\sigma}\,\,
\frac{\left(\epsilon^n_{l\perp}-s^{\prime} \kappa_n  \mu_N  B\right)m^*}
{\displaystyle   \epsilon^n_{l s^{\prime}} \epsilon^n_{l\perp}}
\frac{1 }{\displaystyle 1+ \frac{g_\sigma^2}{{m'}^2_\sigma}\, 
\left(I_1+I_2\right)},
\eeq
with the integrals $I_1$ and $I_2$ having the form given in Eq.(\ref{i1}) and Eq.(\ref{i2}).

\subsection{Current derivatives}
The baryonic currents  Eqs.(\ref{jp})-(\ref{jn}) in the asymmetric nuclear system can be written in terms of the components in directions parallel $\hat{\boldsymbol{k}}$ and perpendicular $\hat{\boldsymbol{n}}_{\perp}$ to the external magnetic field.
Using explicitly the expression for the effective momentum $\boldsymbol{{\cal K}}^j_{i}= \boldsymbol{k}_{i} - \boldsymbol{{\cal V}}^j_{i}, \quad j=p,\, n$ we have, for protons
\beq
\boldsymbol{j}_{p}=j_{pz} \hat{\boldsymbol{k}},
\eeq
with 
\beq
j_{pz} = \sum_{j s^{\prime\prime}\nu^{\prime\prime}} n^p_{j s^{\prime\prime}\nu^{\prime\prime}}
 \frac{\displaystyle k_{zj}-{\cal    V}_{pz}}{\displaystyle \epsilon^p_{j s^{\prime\prime}\nu^{\prime\prime}}},
\eeq
and for neutrons
\beq
\boldsymbol{j}_{n}= \sum_{j s^{\prime\prime}} 
n^n_{j s^{\prime\prime}}
\left[ 
\frac{\displaystyle k_{zj}-{\cal    V}_{nz}}
{\displaystyle  \epsilon^n_{j s^{\prime\prime}}} \hat{\boldsymbol{k}}
+ 
\frac{\displaystyle \left(\epsilon^n_{j\perp}-s^{\prime\prime} \kappa_n  \mu_N  B\right)
\left(k_{j\perp}-{\cal V}_{n \perp}\right)}
{\displaystyle   \epsilon^n_{j s^{\prime\prime}} \epsilon^n_{j\perp}
}\hat{\boldsymbol{n}}_{\perp}
\right].
\eeq

Using the conventions as defined in Matsui's work \cite{matsui} ${\bf j}_B=\bf{j}_p+\bf{j}_n$ and ${\bf j}_3=\bf{j}_p - \bf{j}_n$  the derivatives of the baryonic isovector current components are obtained assuming that the macroscopic currents in equilibrium will vanish, i. e., ${\bf j}_B={\bf 0}$ and ${\bf j}_3={\bf 0}$. In this way the derivative of the baryonic  vector current (z and $\perp$-components) with respect to the proton density $ n^p_{ls}$ are: 
\begin{eqnarray}
 \frac{\partial {\bf j}_B|_z} {\partial   n^p_{ls}}&=&
\left[\frac{k^p_{lz}}{\epsilon^p_{ls}}-\frac{g^2_\rho}{4 m^2_{\rho}} J_1 \frac{\partial {\bf j}_3|_z} {\partial   n^p_{ls}}\right]\left(1+\frac{g^2_\omega}{m^{'2}_{\omega}}J_2\right)^{-1},
\label{jbnpz}
\end{eqnarray}
with  integrals $J_1$ and $J_2$ that can be decomposed as $J_1=b_p - b_n$ and  $J_2=b_p+ b_n$. We also have,
\beq
 \frac{\partial {\bf j_B}|_{\perp}} {\partial   n^p_{ls}}=0.
\eeq
For the derivatives of ${\bf j_B}$ with respect to the neutron density $n^n_{ls}$ we have
\bea
 \frac{\partial  {\bf j}_B|_z} {\partial   n^n_{ls}}&=&
\left[\frac{k^n_{lz}}{\epsilon^n_{ls}}-\frac{g^2_\rho}{4 m^2_{\rho}} J_1 \frac{\partial {\bf j_3}|_z} {\partial   n^n_{ls}}\right]\left(1+\frac{g^2_\omega}{m^{'2}_{\omega}}J_2\right)^{-1},
\label{jbnnz}
\eea
\bea
 \frac{\partial {\bf {j_B}}|_{\perp}} {\partial   n^n_{ls}}&=&
\left[ \frac{(\epsilon^n_{\perp l}-s\kappa_{n} \mu_{N} B) }{\epsilon^n_{\perp l}}\frac{{\boldsymbol{k}^n}|_{\perp}}{\epsilon^n_{i s}}
+\frac{g^2_\rho}{4 m^2_{\rho}}J_3  \frac{\partial {\bf j_3}_{\perp}} {\partial   n^n_{ls}}\right]
\left(1+\frac{g^2_\omega}{m^{'2}_{\omega}}J_3\right)^{-1},
\label{jbnp}
\eea
with the integral $J_3$ which can be written out as  $J_3= I_3$. If we now calculate the derivatives of the ${\mathbf j}_3$ current with respect to $n^p_{ls}$
\bea
 \frac{\partial {\bf j_3}|_z} {\partial   n^p_{ls}}&=&
\left[\frac{k^p_z}{\epsilon^p_{s}}-\frac{g^2_\omega}{m^{'2}_{\omega}}  J_1 \frac{\partial 
{{\bf j_B}|_z}} {\partial   n^p_{ls}}\right] \left[1+\frac{g^2_\rho}{4 m^2_{\rho}}J_2\right]^{-1},
\label{j3npz}
\eea

\beq
 \frac{\partial {\bf j_3}|_{\perp}} {\partial   n^p_{ls}}=0,
\eeq
and with respect to neutron density $n^n_{ls}$,
\bea
 \frac{\partial {\bf j_3}|_z} {\partial   n^n_{ls}}&=&
\left[\frac{-k^n_{lz}}{\epsilon^n_{ls}}-\frac{g^2_\omega}{m^{'2}_{\omega}} J_1 \frac{\partial {\bf j_B}|_z} {\partial   n^n_{ls}}\right] \left[1+\frac{g^2_\rho}{4 m^2_{\rho}}J_2 \right]^{-1},
\label{j3nnz}
\eea
\bea
 \frac{\partial {\bf j_3}|_{\perp}} {\partial   n^n_{ls}}&=&
\left[ \frac{-(\epsilon^n_{\perp l}-s\mu_{N}\kappa_{n}B) }{\epsilon^n_{\perp l}}\frac{{\boldsymbol {k}^n}|_{\perp}}{\epsilon^n_{l s}}
+\frac{g^2_\omega}{m^{'2}_{\omega}}J_3  \frac{\partial {\bf j_B}|_{\perp}} {\partial   n^n_{ls}}\right]
\left[1+\frac{g^2_\rho}{4 m^2_{\rho}}J_3\right]^{-1}.
\label{j3nnp}
\eea

To solve the coupled set of Eqs.(\ref{jbnpz})-(\ref{j3nnp}) for the derivatives of ${\bf j}_B$ and ${\bf j}_3$ the following formula can be used in a straightforward way. For the system,
\beq
y_1= \frac{a+b y_2} {1+d}, \, \,y_2= \frac{c+e y_1}{1+f} \label {y1y2},
\eeq
the solutions are:
\bea
y_1=\frac{a (1+f)+bc}{(1+d)(1+f)-b e},\,\, y_2=\frac{c (1+d)+ea}{(1+d)(1+f)-b e}.
\label{soly1}
\eea

\begin{acknowledgments}

M. A. P. G. acknowledges kind hospitality of University of Coimbra, where part of this work was developed, she also acknowledges useful comments from C. Albertus and S. Marcos. This work was partially supported by FEDER and FCT (Portugal) under the projects CERN/FP/109316/2009  and  PTDC/FP/64707/2006, Spanish Ministry of Educacion under project FIS-2009-07238 and by the  University of Salamanca-University of Coimbra treaty of Collaboration and COMPSTAR.

\end{acknowledgments}

\end{document}